\documentclass[11pt,a4paper]{article}
\pdfoutput=1
\usepackage{jheppub}
\usepackage{amsfonts}
\usepackage{bbm}
\usepackage{graphicx}
\usepackage{caption}
\usepackage{subcaption}
\usepackage{bigints}
\usepackage[normalem]{ulem}
\usepackage{slashed}

\newcommand{\RN}[1]{%
  \textup{\uppercase\expandafter{\romannumeral#1}}%
}

\def\bea{\begin{eqnarray}}
\def\eea{\end{eqnarray}}
\def\be{\begin{equation}}
\def\ee{\end{equation}}
\def\ba{\begin{align}}
\def\ea{\end{align}}

\newcommand{\bem}{\begin{pmatrix}}
\newcommand{\eem}{\end{pmatrix}}

\def\sl{$SL(2,R)$}

\def\={\;  = \;}
\def\+{\, + \,}

\def\rt2{\sqrt{2}}

\title{Pole skipping and chaos in anisotropic plasma: a holographic study}

\author[a,b]{\small{Karunava Sil}}
\affiliation[a]{\small{School of Basic Sciences, Indian Institute of Technology Bhubaneswar, Bhubaneswar 752050, India}}
\affiliation[b]{\small{Department of Physics, Indian Institute of Technology Ropar,
Rupnagar, Punjab 140001, India}}

\emailAdd{ks45@iitbbs.ac.in}

\abstract{Recently, a direct signature of chaos in many body system has been realized from the energy density retarded Green's function using the phenomenon of `pole skipping'. Moreover, special locations in the complex frequency and momentum plane are found, known as the pole skipping points such that the retarded Green's function can not be defined uniquely there. In this paper, we compute the correction/shift to the pole skipping points due to a spatial anisotropy in a holographic system by performing near horizon analysis of EOMs involving different bulk field perturbations, namely the scalar, the axion and the metric field. For vector and scalar modes of metric perturbations we construct the gauge invariant variable in order to obtain the master equation. Two separate cases for every bulk field EOMs is considered with the fluctuation propagating parallel and perpendicular to the direction of anisotropy. We compute the dispersion relation for momentum diffusion along the transverse direction in the shear channel and show that it passes through the first three successive pole skipping points. The pole skipping phenomenon in the sound channel is found to occur in the upper half plane such that the parameters Lyapunov exponent $\lambda_{L}$ and the butterfly velocity $v_{B}$ are explicitly obtained thus establishing the connection with many body chaos.}

\makeatletter
\gdef\@fpheader{}
\makeatother

\begin{document}

%

\maketitle

\section{Introduction}
A large number of studies has been conducted in recent years to quantify the chaotic behavior of quantum systems with large number of degrees of freedom. Classically, the chaotic behavior of a dynamical system is characterized by a parameter $\lambda_{L}$, known as the Lyapunov exponent. A positive value of $\lambda_{L}$ indicates an exponentially fast growth (w.r.t time $t$) of separation between two phase space trajectories which were infinitesimally close at some initial time $t_{0}$. In quantum system, the characteristics of chaotic behavior is somewhat analogous to that of the classical systems. However, the mathematical tools required to diagnose or quantitatively compute the chaotic behavior in quantum systems are different from the classical ones. For instance, in quantum systems the random matrix theory is one such commonly used tools to describe chaos \cite{Ullmo}. However, due to the progress in the last few years it turns out that a more satisfactory diagnosis of quantum chaos can be achieved from the study of black hole, in particular using the holographic principle. \footnote {See \cite{Jahnke:2018off} for a complete review of recent work on holographic chaos.}

The gauge/gravity duality and for that matter, the AdS/CFT correspondence \cite{Maldacena:1997re, Witten:1998qj, Aharony:1999ti} provides a significant improvement of our understanding on quantum systems with large number of degrees of freedom at strong coupling. Using the tools of gauge/gravity duality one can perform the gravitational shock wave analysis \cite{Shenker:2013pqa, Roberts:2014isa, Shenker:2013yza, Shenker:2014cwa, kitaev} to calculate the out-of-time ordered correlation function (OTOC) which is regarded as the measure of chaos in quantum systems.
The OTOC characterises the chaotic behavior in many body quantum systems in terms of two parameters, namely the lyapunov exponent $\lambda_{L}$ and the butterfly velocity $v_{B}$. A series of comprehensive work has already been done over the past few years in order to establish this connection between OTOC and quantum chaos, for instance see \cite{Shenker:2013pqa, Roberts:2014isa, Shenker:2014cwa, Larkin,  Maldacena:2015waa, Kitaev:2017awl, Polchinski:2015cea, Jensen:2016pah, Hashimoto:2017oit} and the references therein. More precisely, in chaotic systems the OTOC or essentially the four point correlation function shows the following exponential growth w.r.t time and space,
\begin{equation}\label{OTOC}
\langle V(t,\vec{x})W(0)V(t,\vec{x})W(0)\rangle_{\beta_{0}}\simeq 1-e^{\lambda_{L}(t-|\vec{x}|/v_{B})},
\end{equation}
$V$ and $W$ being some generic operator. The butterfly velocity $v_{B}$ represents the speed at which the perturbation propagates in space. However, most recently it is observed that a particular component of a much more simple correlation function namely, the energy density retarded Green's function can provide a direct signature of quantum chaos in most of the holographic theories with Einstein gravity \cite{Blake:2018leo, Blake:2019otz, Natsuume:2019xcy, Natsuume:2019sfp, Natsuume:2020snz}.

In strongly coupled quantum field theories at finite temperature, thermal retarded Green's function encode information about the near equilibrium physics of the system. Using the concept of gauge/gravity duality the properties of these thermal Green's functions has been investigated for strongly coupled systems in \cite{Son:2002sd, Gubser:1998bc, Herzog:2002pc, Skenderis:2008dh}. It is recently observed that the dispersion relation of collective excitations in the energy density Green's function is actually related to the particular form of the OTOC as given in (\ref{OTOC}). In particular, the parameters $\lambda_{L}$ and $v_{B}$ can be obtained by analysing the behavior of quasinormal modes in holographic systems. The phenomenon that sets up a direct relation between these collective excitations in the retarded Green's function and the parameters of chaos in quantum many body systems is known as the `pole skipping' phenomenon. It is defined as the special locations in the complex $(w-q)$ plane such that the lines of zeroes and the lines of poles of the retarded Green's function in momentum space coincide or crosse each other at that special point and hence it is not defined uniquely.

The non-uniqueness of Green's function at special location in the complex plane was explicitly shown for example in \cite{Blake:2019otz, Natsuume:2019xcy} by performing a near horizon expansion of the equation of motion. To explain the basic idea with a particular example, let us consider the equation of motion for a massive scalar field $\phi$ in planner AdS black hole background with horizon radius $r_{H}$ \cite{Blake:2019otz}. Writing the metric in terms of the ingoing Eddington-Finkelstein coordinates $(v,r,x)$ with $\phi=\varphi(r)e^{-iwv+iqx}$, one obtains the second order equation of motion for $\varphi(r)$. Now for the near horizon behavior of the solution, $\varphi(r)$ can be expanded as $\varphi(r)=(r-r_{H})^{\lambda}$ with the two possible results for $\lambda$ as $\lambda_{1}=0$, $\lambda_{2}=(iw/2\pi T)$, $T$ being the black hole temperature. At this point one imposes the ingoing wave boundary condition at the horizon and picks the exponent $\lambda_{1}$ which makes the solution regular at the horizon. However at special values of $w_{n}=-i2\pi T n$, $(n=1,2,3,..)$, the other exponent $\lambda_{2}$ becomes a positive integer and hence there exist two independent ingoing regular solutions at the horizon. As a result the retarded Green's function can not be defined uniquely. In fact it turns out that the retarded Green's function actually depends on the slope with which one approaches the pole skipping points.

In order to have a clear idea about the phenomenon, it would be helpful to describe in short the procedure to obtain the pole skipping points at different orders of expansion near the horizon. As already discussed, the pole skipping points are defined as the special locations in the complex $(w-q)$ plane at which the pole of the retarded Green's function is skipped because the numerator as well as the denominator (of the Green's function) vanishes simultaneously. These special points are represented by the set of values $(w_{n},q_{n})$ obtained by the near horizon analysis of the equation of motion involving the bulk fields in the dual gravity theory. In the above the index $n$ can take positive integer values $(n=0,1,2,3,..)$ that indicates the order for the near horizon expansion of the bulk equation.
Let us consider an equation of motion for any bulk field $Z(r)$ having the following general form,
\begin{equation}\label{geqn}
Z^{\prime\prime}+A(r)Z^{\prime}+B(r)Z=0,
\end{equation}
where $r$ denotes the radial coordinate of the dual gravitational background and the horizon is defined at $r=r_{H}$. In this analysis we will be using the ingoing Eddington-Finkelstein coordinates in which the metric for a generic gravity background takes the following form,
\begin{equation}
ds^2=-f(r)dv^2+2dv dr+...
\end{equation}
where $v$ is defined in terms of the tortoise coordinate $r_{*}$ as $v=t+r_{*}$ and the ellipses indicates the spatial part of the metric. To proceed further, consider the near horizon expansion of the bulk field $Z(r)$ as,
\begin{equation}\label{pex}
Z(r)=\sum_{n}z_{n}\left(r-r_{H}\right)^n
\end{equation}
and then put it back into the equation of motion (\ref{geqn}). The resultant equation can be expanded in a power series near the horizon $r_{H}$ and are given at different orders as,
\begin{equation}\label{eqnorder}
\begin{split}
& c_{00}z_{0}+c_{01}z_{1}=0,\\
& c_{10}z_{0}+c_{11}z_{1}+c_{12}z_{2}=0,\\
& c_{20}z_{0}+c_{21}z_{1}+c_{22}z_{2}+c_{23}z_{3}=0,\\
& ~~~~~~~~~~~~~~~~ \vdots
\end{split}
\end{equation}
Considering equations upto $n$th order one can construct a $n\times n$ matrix with elements as the coefficients of the above set of equations is given as,
\begin{equation}
\label{Cmatrix}
C= \begin{pmatrix}
	c_{00} && c_{01} && 0 && 0 && \cdots\\
	c_{10} && c_{11} && c_{12} && 0 && \cdots\\
	c_{20} && c_{21} && c_{22} && c_{23} && \cdots\\
	\vdots && \vdots && \vdots && \vdots && \cdots \\,
\end{pmatrix}
\end{equation}
where the above matrix elements are functions of $w$ and $q$. With the above matrix, the pole skipping points are determined by the solutions of the following equations,
\begin{equation}\label{matrixeqn}
c_{n-1~n}=0,~~~~~~\det{C}=0.
\end{equation}
In this work, we manage to solve the above equation analytically only for the first few pole skipping points and one has to rely on numerical solutions for the results at higher orders.

Several interesting aspects of quantum chaos can be realised from the phenomenon of pole skipping. The connection between hydrodynamics and chaos has been explicitly shown in several holographic theories using the pole skipping. In \cite{Grozdanov:2017ajz}, the authors showed through numerical calculation that for the hydrodynamic sound mode, the dispersion relation provides the results of lyapunov exponent, a parameter for quantum chaos. This connection with hydrodynamics was shown to be valid even for gravitational theories with curvature squared correction in \cite{Grozdanov:2018kkt} using the pole skipping phenomenon. Also in \cite{Wu:2019esr} the author obtained the higher curvature correction to the special value of the momentum at which the pole is supposed to be skipped, with no correction to the results for the frequency. Another interesting connection between the transport coefficient and the parameter of quantum chaos was developed in \cite{Blake:2016wvh, Blake:2016sud}. Moreover, the diffusion constant for both charge and momentum is shown to be related to the square of the butterfly velocity in strongly coupled theory. In holographic theories, both transport coefficients and the parameters of chaos are related to the near horizon physics. So the connection can be realized from the AdS/CFT correspondence.
A list of several other works recently done on pole skipping phenomenon can be found in \cite{Grozdanov:2020koi, Grozdanov:2019uhi, Liu:2020yaf, Ahn:2020bks, Abbasi:2020ykq, Choi:2020tdj, Abbasi:2019rhy, Ahn:2020baf}.\footnote{Pole skipping in two dimensional CFT and also two dimensional BCFT was studied in \cite{Das:2019tga}}

In this paper we have considered a gravitational background with a spatial anisotropy as obtained in \cite{Mateos:2011ix, Mateos:2011tv}.\footnote{see \cite{Chakrabortty:2013kra} for holographic study of brownian motion in the same background} The primary goal of this paper is to find explicitly the corrections that the pole skipping points receive in the complex frequency-momentum plane due to the spatial anisotropy in the background theory parameterized by $a$ (or a dimensionless one $b=a/T$). We consider the bulk scalar, axion and metric field perturbations and in each case we make two different choices for the direction of propagation for the field fluctuations, namely (i) along the direction of anisotropy (ii) perpendicular to the direction of anisotropy. We find that the frequency at the pole skipping point receives no correction due to the anisotropy but only the momentum gets corrected. The rest of the paper is organized as follows, In section-{\bf 2}, we present a short discussion on the gravitational background dual to a SYM plasma with spatial anisotropy. In section-{\bf 3}, we describe quantitatively the pole skipping phenomenon in energy density Green's function in the upper half complex plane by doing a near horizon analysis of the $vv$ component of Einstein's equation and obtain the Lyapunov exponent and butterfly velocity associated to the phenomenon of chaos. In section {\bf 4}, we first study the pole skipping in the lower half plane for scalar and axion field and then carry on the analysis for the metric perturbations. For the metric perturbations we take into account all the non-zero components for the shear and the sound channel and construct the corresponding master equations involving the gauge invariant variables (we present a detailed discussions on the construction of gauge invariant variables for different metric field perturbations in appendix-{\bf A}). In the same section we also solve the dispersion relation of the transverse momentum diffusion in the shear channel using numerical method and showed that it passes through the corresponding pole skipping points. Finally, we conclude in section-{\bf 5}.
\section{Details of the anisotropic background}
In this section we will briefly describe the supergravity solution as obtained by the authors in \cite{Mateos:2011ix, Mateos:2011tv} which is dual to a spatially anisotropic strongly coupled SYM theory at finite temperature. In relativistic heavy ion collision the plasma that is created has been found to be locally anisotropic for a very short time period due to the pressure difference along the longitudinal and transverse direction. This motivates the authors towards a dual gravitational background with anisotropy along a spatial direction.
The five dimensional action involving the metric ($g$), the dilaton ($\phi$) and the axion ($\chi$) field excitation is given as,
\begin{equation}\label{action}
S=\frac{1}{2k^2}\int d^{5}x{\sqrt{-g}\left(R+12-\frac{1}{2}(\partial \phi)^2-\frac{1}{2}e^{2 \phi}(\partial \chi)^2\right)},
\end{equation}
with $2k^2=16\pi G$ as the five dimensional gravitational constant.
The supergravity solution is given by the following five dimensional metric as \cite{Mateos:2011ix, Mateos:2011tv},
\begin{equation}\label{metric1}
ds^2=e^{-\frac{\phi(r)}{2}}r^2\biggl(-\mathcal{F}\mathcal{B}dt^2+\mathcal{H}dx_{1}^2+dx_{2}^2+dx_{3}^2+\frac{dr^2}{r^4\mathcal{F}}\biggl),
\end{equation}
\begin{equation}\label{axidila}
\chi=a x_{1},~~~~\phi=\phi(r)
\end{equation}
where the spatial anisotropy is considered along the $x_{1}$ direction. The above solution is static, completely regular on the horizon and also asymptotically AdS. Notice that the axion field $\chi$ is linearly proportional to $x_{1}$, the anisotropic direction and the dilaton field $\phi$ depends on the radial coordinate $r$. The black hole horizon is located at $r=r_{H}$ with the boundary at $r=\infty$. The explicit form of the different metric components in (\ref{metric1}) is given in \cite{Mateos:2011ix, Mateos:2011tv}, where the authors have introduced an anisotropy parameter $a$. Also, assuming the anisotropy to be weak ($a/T\ll 1$, $T$ being the hawking temperature), the authors have kept terms only upto quadratic order in $a$ in the series expanded form of the metric components of (\ref{metric1}). It is important to note that the anisotropy parameter $a$ is a dimensionfull quantity, $[a]=\textit{dim}[\textrm{Length}]$. The hawking temperature is given upto quadratic order in $a$ as,
\begin{equation}\label{temp}
T=\frac{r_{H}}{\pi}+\frac{a^2}{r_{H}}\frac{\left(5\log{(2)}-2\right)}{48\pi}+\mathcal{O}(a^4).
\end{equation}
Varying above action (\ref{action}) with respect to $g_{\mu\nu}$, $\phi$ and $\chi$ one gets the Einstein equations as well as the equations of motion for the scalars as,
\begin{equation}\label{e1}
\begin{split}
R_{\mu\nu}=-4g_{\mu\nu}+\frac{1}{2}\partial_{\mu}\phi\partial_{\nu}\phi+\frac{1}{2}e^{2\phi}\partial_{\mu}\chi\partial_{\nu}\chi,
\end{split}
\end{equation}
\begin{equation}\label{e2}
\begin{split}
\Box \phi=e^{2\phi}\left(\partial\chi\right)^2,
\end{split}
\end{equation}
\begin{equation}\label{e3}
\begin{split}
\Box\chi =0.
\end{split}
\end{equation}
However for the near horizon analysis we need to write the above five dimensional metric in terms of the ingoing Eddington-Finkelstein coordinates defined as,
\begin{equation}
v=t+r_{*},~~~~r_{*}=\int \left(\frac{g_{rr}}{g_{tt}}\right)^{1/2}dr.
\end{equation}
The metric (\ref{metric1}) can be rewritten in Eddington-Finkelstein coordinate as,
\begin{equation}\label{metric2}
ds^2=g_{vv}dv^2+g_{11}dx_{1}^2+g_{22}\left(dx_{2}^2+dx_{3}^2\right)+2g_{vr}dvdr,
\end{equation}
where the metric components including the correction due anisotropy are given as,
\begin{equation}\label{metriccomp}
\begin{split}
g_{vv}&=-r^2 \left(1-\frac{r_{H}^4}{r^4}\right)+\frac{a^2}{12}\left[1+\frac{r_H^2}{r^2}\left(5~ \log~{2}-1\right)
-\frac{5r^2}{r_{H}^2}\left(1+\frac{r^4}{r_{H}^4}\right) \log
\left(1+\frac{r_H^2}{r^2}\right)\right]\\
g_{vr}&=1-\frac{a^2}{48} \left[\frac{10}{r^2+r_H^2}-\frac{5}{r_H^2}\log
\left(\frac{r_H^2}{r^2}+1\right)\right]\\
g_{11}&=r^2+\frac{3a^2}{8}\left[\frac{r^2}{r_H^2}\log \left(1+\frac{r_H^2}{r^2}\right)\right]\\
g_{22}&=r^2+\frac{a^2}{8}\left[\frac{r^2}{r_H^2}\log \left(1+\frac{r_H^2}{r^2}\right)\right].
\end{split}
\end{equation}
Using the above metric, in the following sections we will do the near horizon analysis of the EOM for different field perturbation to obtain the special points in the complex $(w-q)$ plane where the pole skipping phenomenon will be explicit.
\section{Pole skipping in energy density Green's function}
Recently it is explicitly shown that an universal description of the chaotic behavior in many body system can be achieved by a hydrodynamical effective field theory \cite{Blake:2017ris}.\footnote{Also see \cite{Haehl:2018izb}} In other words, this effective field theory predicts that the exponential growth of the OTOC can be realized from the energy density retarded Green's function in a sense that it exhibit pole skipping at a particular value of frequency and momentum that is directly related to the parameters $\lambda_{L}, v_{B}$ appearing in the exponential form of the OTOC as,
\begin{equation}\label{lyabut}
w=i\lambda_{L},~~~~~~~q=\frac{i\lambda_{L}}{v_{B}}
\end{equation}
So one can obtain $\lambda_{L}, v_{B}$ from the energy density Green's function using the pole skipping phenomenon. In this section we will try to obtain the explicit form of $\lambda_{L}$ and $v_{B}$ from the near horizon expansion of Einstein's equation. In particular we would be interested in the corrections that these parameters receives due to the spatial anisotropy in the background theory. In turns out that only the $vv$ component of Einstein's equation has to be computed near the horizon in order to determine the pole skipping point.

We will start by considering the small perturbation of the above unperturbed metric (\ref{metric2}). There are two different choices that one can make regarding the direction of propagation for the metric perturbation, (i) Perturbation along the direction of anisotropy, (ii) perturbation perpendicular to the direction of anisotropy. In the following we will consider these two cases separately to study the phenomenon of pole skipping for the metric fluctuation in the sound channel.
Let us now consider the following linear perturbations of the above fields as,
\begin{equation}\label{allpert}
\begin{split}
g_{\mu\nu}&=g^{(0)}_{\mu\nu}+h_{\mu\nu},\\
\phi &=\phi_{0}+\varphi,\\
\chi &=\chi_{0}+\psi,
\end{split}
\end{equation}
where, $g^{(0)}_{\mu\nu}, \phi_{0}, \chi_{0}$ represents the background values of the fields and $h_{\mu\nu}, \varphi, \psi$ are their linear perturbations. For computational simplification here we will work in radial gauge such that all the components of metric perturbation which are of the form $h_{r\mu}$ are zero for all $\mu$.
Substituting the above linear fluctuation of the bulk fields to the corresponding equations of motion as given in (\ref{e1}, \ref{e2}, \ref{e3}), we obtained the following linearized equations as,
\begin{equation}\label{lineareqn}
\begin{split}
& R_{\mu\nu}^{(1)}+4h_{\mu\nu}-\frac{1}{2}\left(\partial_{\mu}\phi_{0}\partial_{\nu}\varphi+\partial_{\mu}\varphi\partial_{\nu}\phi_{0}\right)
-e^{2\phi_{0}}\left[\left(\partial_{\mu}\chi_{0}\partial_{\nu}\chi_{0}\right)\varphi
+\frac{1}{2}\left(\partial_{\mu}\psi\partial_{\nu}\chi_{0}+\partial_{\mu}\chi_{0}\partial_{\nu}\psi\right)\right]=0,
\end{split}
\end{equation}
\begin{equation}\label{lineareqn1}
\begin{split}
& g^{(0)\mu\nu}\left[\partial_{\mu}\partial_{\nu}\varphi-2e^{2\phi_{0}}\left(\partial_{\mu}\chi_{0}\partial_{\nu}\chi_{0}\right)\varphi
+\left(\partial_{\mu}\chi_{0}\partial_{\nu}\psi+\partial_{\mu}\psi\partial_{\nu}\chi_{0}\right)
-\Gamma^{(1)\rho}_{\mu\nu}\partial_{\rho}\phi_{0}
-\Gamma^{(0)\rho}_{\mu\nu}\partial_{\rho}\varphi
\right]\\
&-h^{\mu\nu}\left[\partial_{\mu}\partial_{\nu}\phi_{0}+e^{2\phi_{0}}\left(\partial_{\mu}\chi_{0}\partial_{\nu}\chi_{0}\right)
-\Gamma^{(0)\rho}_{\mu\nu}\partial_{\rho}\phi_{0}\right]=0,
\end{split}
\end{equation}
\begin{equation}\label{lineareqn2}
\begin{split}
& g^{(0)\mu\nu}\left(\partial_{\mu}\partial_{\nu}\psi-\Gamma^{(1)\rho}_{\mu\nu}\partial_{\rho}\chi_{0}
-\Gamma^{(0)\rho}_{\mu\nu}\partial_{\rho}\psi\right)-h^{\mu\nu}\left(\partial_{\mu}\partial_{\nu}\chi_{0}
-\Gamma^{(0)\rho}_{\mu\nu}\partial_{\rho}\chi_{0}\right)=0,
\end{split}
\end{equation}
where, $\Gamma^{(0)\rho}_{\mu\nu}$ is the background value of the affine connection while $\Gamma^{(1)\rho}_{\mu\nu}$, $R^{(1)}_{\mu\nu}$ are the linearized fluctuations to the affine connection and the ricci tensor respectively defined as,
\begin{equation}\label{ricci}
\begin{split}
\Gamma^{(0)\rho}_{\mu\nu}&=\frac{1}{2}g^{(0)\lambda\rho}\left(\partial_{\mu}g^{(0)}_{\lambda\nu}+\partial_{\nu}g^{(0)}_{\lambda\mu}
-\partial_{\lambda}g^{(0)}_{\mu\nu}\right),\\
\Gamma^{(1)\rho}_{\mu\nu}&=\frac{1}{2}\left[g^{(0)\lambda\rho}\left(\partial_{\mu}h_{\lambda\nu}+\partial_{\nu}h_{\lambda\mu}
-\partial_{\lambda}h_{\mu\nu}\right)-h^{\lambda\rho}\left(\partial_{\mu}g^{(0)}_{\lambda\nu}+\partial_{\nu}g^{(0)}_{\lambda\mu}
-\partial_{\lambda}g^{(0)}_{\mu\nu}\right)\right],\\
R^{(1)}_{\mu\nu}&=\partial_{\rho}\Gamma^{(1)\rho}_{\mu\nu}-\partial_{\mu}\Gamma^{(1)\rho}_{\rho\nu}
+\Gamma^{(0)\rho}_{\rho\lambda}\Gamma^{(1)\lambda}_{\mu\nu}+\Gamma^{(1)\rho}_{\rho\lambda}\Gamma^{(0)\lambda}_{\mu\nu}
-\Gamma^{(0)\rho}_{\mu\lambda}\Gamma^{(1)\lambda}_{\rho\nu}-\Gamma^{(1)\rho}_{\mu\lambda}\Gamma^{(0)\lambda}_{\rho\nu}
\end{split}
\end{equation}
\subsection{Perturbation parallel to the direction of anisotropy}
We first consider the perturbations to propagate along the direction of the anisotropy, $x_{1}$ so that one can use the fourier transform to write the same as,
\begin{equation}\begin{split}\label{fluctuation}
h_{\mu\nu}(v,r,x_{1})&=e^{-iwv+iqx_1}h_{\mu\nu}(r),
\\ \varphi(v,r,x_{1})&=e^{-iwv+iqx_1}\varphi_{0}(r),
\\ \psi(v,r,x_{1})&=e^{-iwv+iqx_1}\psi_{0}(r).
\end{split}
\end{equation}
Moreover, with this particular choice of the field fluctuation one can categorize all the metric perturbations into three different modes depending on their transformation under the $SO(2)$ rotational symmetry in the $x_{2}-x_{3}$ plane \cite{Policastro:2002se}, namely (i) Scalar mode, (ii) Vector mode, (iii) Tensor mode. In the following we will write down the nonzero components of these three modes of metric perturbation.
\begin{itemize}\label{parallel}
\item Scalar modes: $h_{vv},h_{vr},h_{rr}, h_{vx_{1}},h_{rx_{1}}, h_{x_{1}x_{1}}, h_{x_{2}x_{2}}=h_{x_{3}x_{3}}$,
\item Vector modes: $h_{vx_{2}},h_{rx_{2}}, h_{x_{1}x_{2}}$,
\item Tensor mode: $h_{x_{2}x_{3}}$.
\end{itemize}
For the computation of special point in the complex $(w-q)$ plane we consider only the sound modes of metric perturbation which corresponds to the retarded Green's function for the temporal component of the energy momentum tensor, $G^{R}_{T^{00}T^{00}}$.

To proceed further we consider the near horizon expansion of the above fluctuations to have the following form,
\begin{equation}\label{nhe}
Y(r)=Y^{(0)}+Y^{(1)}\left(r-r_{H}\right)+.....,
\end{equation}
where $Y$ represents in general the fluctuations of the metric, scalar and the axion field.
The reason for the above near horizon expansion is due to the fact that the location of the special point depends on the near horizon value of the background metric. Substituting (\ref{nhe}) into the linearized Einstein equation one gets the following result for the $vv$ component in the near horizon limit as,
\begin{equation}\label{einsnh}
\begin{split}
&-i\Biggl(\frac{2iw+4r_{H}}{r_{H}^2}-\frac{a^2}{6r_{H}^3}\left(1+\frac{\log(2)}{2}+\frac{3i w\log(2)}{2r_{H}}\right)\Biggr)
\Biggl(2wh_{x_{2}x_{2}}^{(0)}+\left(1-\frac{a^2\log(8)}{8r_{H}^2}\right)(2qh_{vx_{1}}^{(0)}+wh_{x_{1}x_{1}}^{(0)})\Biggr)\\
&+\left(-\frac{6iw}{r_{H}}\left(1+\frac{5a^2\log(2)}{48r_{H}^2}\right)+\frac{2q^2}{r_{H}^2}\left(1-\frac{a^2\log(8)}{8r_{H}^2}\right)
\right)h^{(0)}_{vv}=0.
\end{split}
\end{equation}
The above equation is identically satisfied for the particular value of $w$ and $q$,
\begin{equation}
w=2i\pi T,~~~~~q=i\sqrt{6}\pi T\left[1+b^2\left(\frac{1+4\log(2)}{48\pi^2}\right)\right],
\end{equation}
where, we define the dimensionless quantity $b$ defined as $b=a/T\ll 1$, in the limit $a\ll T$. The Lyapunov exponent and the butterfly velocity can be obtained as (\ref{lyabut}),
\begin{equation}\label{butt1}
\lambda_{L}=2\pi T,~~~~~v^2=\frac{|w|^2}{|q|^2}=\frac{2}{3}-b^2\left(\frac{1+4\log(2)}{36\pi^2}\right).
\end{equation}
The Lyapunov exponent takes the maximum value allowed by the chaos bound even in the presence of a spatial anisotropy and only the butterfly velocity receives a correction due to the anisotropy.
\subsection{Perturbation perpendicular to the direction of anisotropy}
We now consider the perturbation along the $x_{2}$ coordinate, that is perpendicular to the direction of anisotropy. So the perturbation of the fields in this case can be written as,
\begin{equation}\begin{split}\label{field}
h_{\mu\nu}(v,r,x_{2})&=e^{-iwv+iqx_2}h_{\mu\nu}(r),
\\ \varphi(v,r,x_{2})&=e^{-iwv+iqx_2}\varphi_{0}(r),
\\ \psi(v,r,x_{2})&=e^{-iwv+iqx_2}\psi_{0}(r).
\end{split}
\end{equation}
In this case the non zero components of the metric perturbations for the scalar, vector and the tensor mode are given as,
\begin{itemize}\label{perp}
\item Scalar modes: $h_{vv},h_{vr},h_{rr}, h_{vx_{2}},h_{rx_{2}}, h_{x_{2}x_{2}}, h_{x_{1}x_{1}}=(g_{11}/g_{22})h_{x_{3}x_{3}}$,
\item Vector modes: $h_{vx_{3}},h_{rx_{3}}, h_{x_{2}x_{3}}$,
\item Tensor mode: $h_{x_{1}x_{3}}$.
\end{itemize}
Again analysing the $vv$ component of the linearized Einsteins equation near the horizon one gets,
\begin{equation}\label{einsnh22}
\begin{split}
&-i\Biggl(\frac{2iw+4r_{H}}{r_{H}^2}-\frac{a^2}{6r_{H}^3}\left(1+\frac{13\log(2)}{2}+\frac{9i w\log(2)}{2r_{H}}\right)\Biggr)
\Biggl(2w h_{x_{1}x_{1}}^{(0)}+\left(1+\frac{a^2\log(2)}{4r_{H}^2}\right)(2qh_{vx_{2}}^{(0)}+wh_{x_{2}x_{2}}^{(0)})\Biggr)\\
&+\left\{-\frac{6iw}{r_{H}}\left(1-\frac{5a^2\log(2)}{48r_{H}^2}\right)+\frac{2q^2}{r_{H}^2}
\left(1-\frac{a^2\log(2)}{8r_{H}^2}\right)\right\}h^{(0)}_{vv}=0.
\end{split}
\end{equation}
The corresponding value of $w$ and $q$ from the above equation can be obtained as,
\begin{equation}\label{gtgt}
w=2i\pi T,~~~~~q=i\sqrt{6}\pi T\left[1-b^2\left(\frac{-1+2\log(2)}{48\pi^2}\right)\right],
\end{equation}
with the butterfly velocity given as,
\begin{equation}\label{butt2}
\lambda_{L}=2\pi T,~~~~~v^2=\frac{|w|^2}{|q|^2}=\frac{2}{3}+b^2\left(\frac{-1+2\log(2)}{36\pi^2}\right)
\end{equation}
Similar to the previous case, here the Lyapunov exponent remains the same and the butterfly velocity gets corrected. The results for the butterfly velocity as obtained here in (\ref{butt1}) and (\ref{butt2}) matches exactly with the results obtained in \cite{Jahnke:2017iwi} using gravitational shock wave analysis.
\section{Gauge invariant variable and pole skipping phenomenon in anisotropic plasma}
In the previous section we have considered the near horizon analysis of only the $(vv)$-component of the Einsteins equation (\ref{lineareqn}) in the sound channel to figure out the location of the lowest order pole skipping point and from that we also obtain the Lyapunov exponent and the  butterfly velocity. However even more rigorous way of doing the same is to take into account all components of the Einsteins equation and study their behavior near the event horizon. For example, in the sound channel there are different components of the metric fluctuation and hence we required to solve multiple equations simultaneously. In particular, we will construct a gauge invariant master variable for the anisotropic background so that all the components of the Einstein's equation can be put together into a single equation which is easier to deal with. In the following, we will first discuss the computation of pole skipping points at different orders for the scalar and the axion field and then we will move on to the fluctuations of the metric corresponding to both the shear and the sound channel where the gauge invariant master variable will play an important role.
\subsection{Scalar field fluctuation}
The equation of motion for the scalar field follows from (\ref{e2}) with the background metric components in terms of Eddington-Finkelstein coordinates as given in (\ref{metriccomp}). As before we have considered two separate choices for the field fluctuation to propagate along the direction of anisotropy or perpendicular to that.
\begin{itemize}
\item{\it Parallel case}
\end{itemize}
Inserting the form of the scalar field perturbation as given in (\ref{fluctuation}), propagating along $x_1$ into (\ref{e2}) one gets equation of motion for the scalar field as,
\begin{equation}\label{scleq}
\varphi_{0}^{\prime\prime}-\biggl(S_{1}+a^2\tilde{S_{1}}\biggr)\varphi_{0}^{\prime}-\biggl(S_{2}+a^2\tilde{S_{2}}\biggr)\varphi_{0}=0.
\end{equation}
where the coefficients $S_1, S_2, \tilde{S_{1}}$ and $\tilde{S_{2}}$ are given in Appendix-{\bf B}. Using the near horizon power series expansion of $\varphi_{0}$ as in (\ref{pex}) one can construct the coefficient matrix $C$ as in (\ref{Cmatrix}). The first few elements of the same is given below as,
\begin{equation}\label{ccomp1}
\begin{split}
& c_{00}=\frac{1}{64r_{H}^5}\biggl[-16r_{H}^2\left(q^2+3iwr_{H}\right)+a^2\biggl(iwr_{H}\left(3+5\log 2\right)-32r_{H}^2+q^2\left(1+6\log 2\right)\biggr)\biggr]\\
& c_{01}=\frac{1}{96r_{H}^3}\biggl[96r_{H}^3-48iwr_{H}^2+a^2\biggl(iw\left(-2+5\log2\right)\biggr)\biggr]\\
& c_{10}=\frac{1}{128r_{H}^6}\biggl[48r_{H}^2\left(q^2+iwr_{H}\right)-a^2\biggl(iwr_{H}\left(4+5\log2\right)-96r_{H}^2+6q^2\left(2+3\log2\right)\biggr)\biggr]\\
& c_{11}=\frac{1}{192r_{H}^5}\biggl[-48r_{H}^2\left\{q^2+2r_{H}\left(2iw-5r_{H}\right)\right\}
+a^2\biggl(4iwr_{H}\left(4+5\log2\right)-138r_{H}^2+3q^2\left(1+6\log2\right)\biggr)\biggr]\\
& c_{12}=\frac{1}{48r_{H}^3}\biggl[192r_{H}^3-48iwr_{H}^2+a^2\biggl(iw\left(-2+5\log2\right)\biggr)\biggr]
\end{split}
\end{equation}
Solving (\ref{matrixeqn}), the location of the first and the second order special points can be obtained analytically and they are given as
\begin{equation}\begin{split}
&w_{1}=-i2\pi T,~~~~q_{1}^2=-6r_{H}^2-\frac{a^2}{4}\biggl(7+9\log{2}\biggr),\\&
w_{2}=-i4\pi T,~~~~~q_{2}^2=\biggl(-12\pm4\sqrt{3}\biggr)r_{H}^2-\frac{a^2}{2}\biggl(2\pm\sqrt{3}+3\left(3\mp\sqrt{3}\right)\log 2\biggr)
\end{split}
\end{equation}
The pole skipping points appearing at higher orders can be calculated numerically for a given value of the dimensionless parameter $b$.
\begin{itemize}
\item{\it Perpendicular case}
\end{itemize}
Considering the perturbation propagating along $x_{2}$ direction we get the following equation of motion for the scalar field,
\begin{equation}\label{scleq2}
\varphi_{0}^{\prime\prime}-\biggl(S_{1}+a^2\tilde{\tilde{S_{1}}}\biggr)\varphi_{0}^{\prime}-\biggl(S_{2}+a^2\tilde{\tilde{S_{2}}}\biggr)\varphi_{0}=0,
\end{equation}
where the coefficients $\tilde{\tilde{S_{1}}}$ and $\tilde{\tilde{S_{2}}}$ are given in Appendix-{\bf A}. In this case the first few elements of the coefficient matrix $C$ are given as
\begin{equation}\label{ccomp2}
\begin{split}
& c_{00}=\frac{1}{192r_{H}^5}\biggl[-48r_{H}^2\left(q^2+3iwr_{H}\right)+a^2\biggl(3iwr_{H}\left(3+5\log (2)\right)-96r_{H}^2+q^2\left(3+6\log ( 2)\right)\biggr)\biggr]\\
& c_{01}=\frac{1}{96r_{H}^3}\biggl[96r_{H}^3-48iwr_{H}^2+a^2\biggl(iw\left(-2+5\log(2)\right)\biggr)\biggr]\\
& c_{10}=\frac{1}{128r_{H}^6}\biggl[48r_{H}^2\left(q^2+iwr_{H}\right)-a^2\biggl(iwr_{H}\left(4+5\log(2)\right)-96r_{H}^2+2q^2\left(2+3\log2\right)\biggr)\biggr]\\
& c_{11}=\frac{1}{192r_{H}^5}\biggl[-48r_{H}^2\left\{q^2+2r_{H}\left(2iw-5r_{H}\right)\right\}
+a^2\biggl(4iwr_{H}\left(4+5\log(2)\right)-138r_{H}^2+3q^2\left(1+2\log(2)\right)\biggr)\biggr]\\
& c_{12}=\frac{1}{48r_{H}^3}\biggl[192r_{H}^3-48iwr_{H}^2+a^2\biggl(iw\left(-2+5\log(2)\right)\biggr)\biggr]
\end{split}
\end{equation}
Again only the first and the second ordered pole skipping points can be solved analytically and they are given below as,
\begin{equation}\begin{split}
&w_{1}=-i2\pi T,~~~~q_{1}^2=-6r_{H}^2-\frac{a^2}{4}\biggl(7+3\log{2}\biggr),\\&
w_{2}=-i4\pi T,~~~~~q_{2}^2=\biggl(-12\pm4\sqrt{3}\biggr)r_{H}^2-\frac{a^2}{2}\biggl(3+\left(3\mp\sqrt{3}\right)\log 2\biggr)
\end{split}
\end{equation}
\begin{figure}[h]
\centering
\begin{tabular}{ccc}
a)&&b)\\
\includegraphics[width=.5\textwidth]{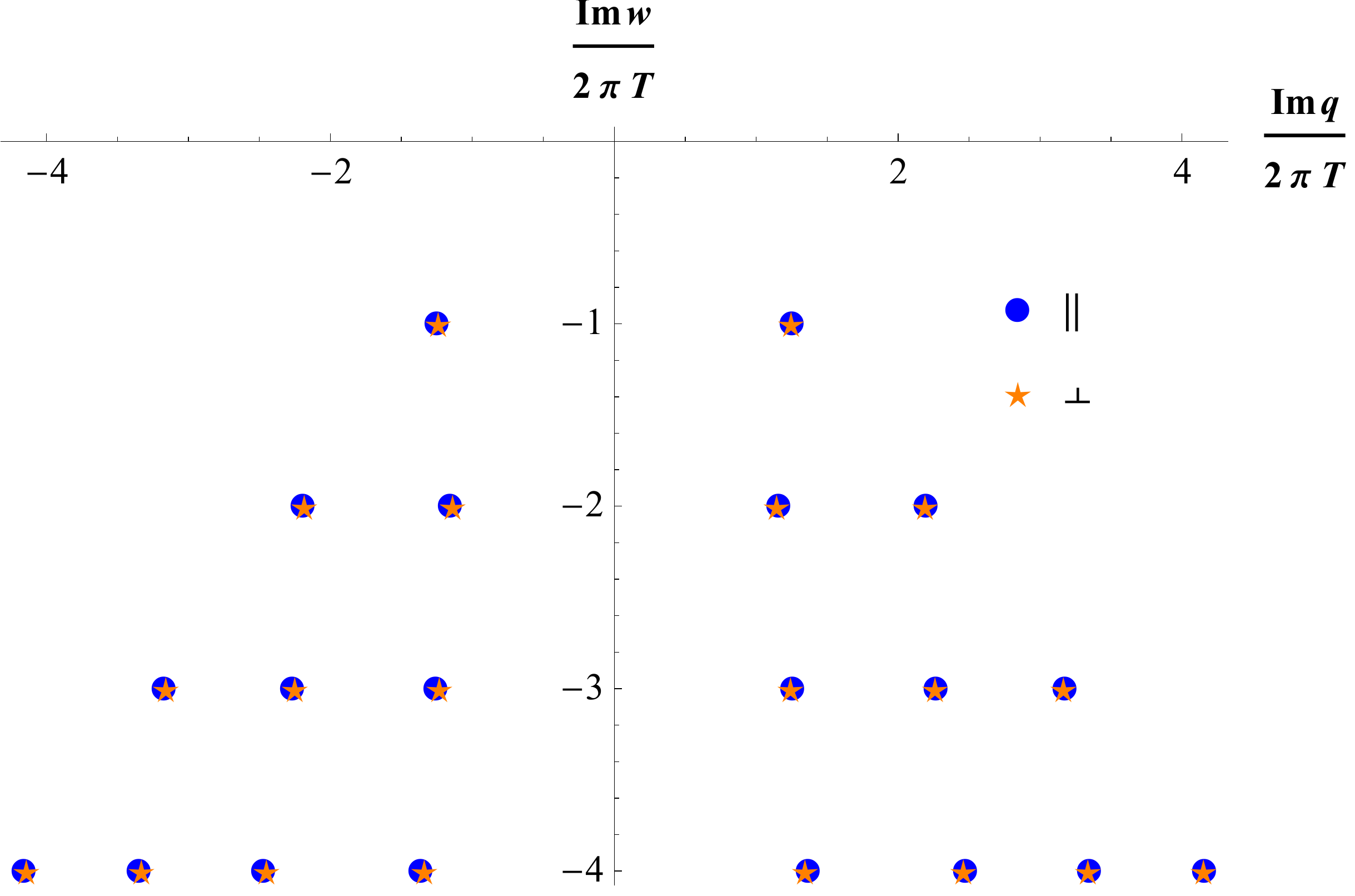}&&
\includegraphics[width=.5\textwidth]{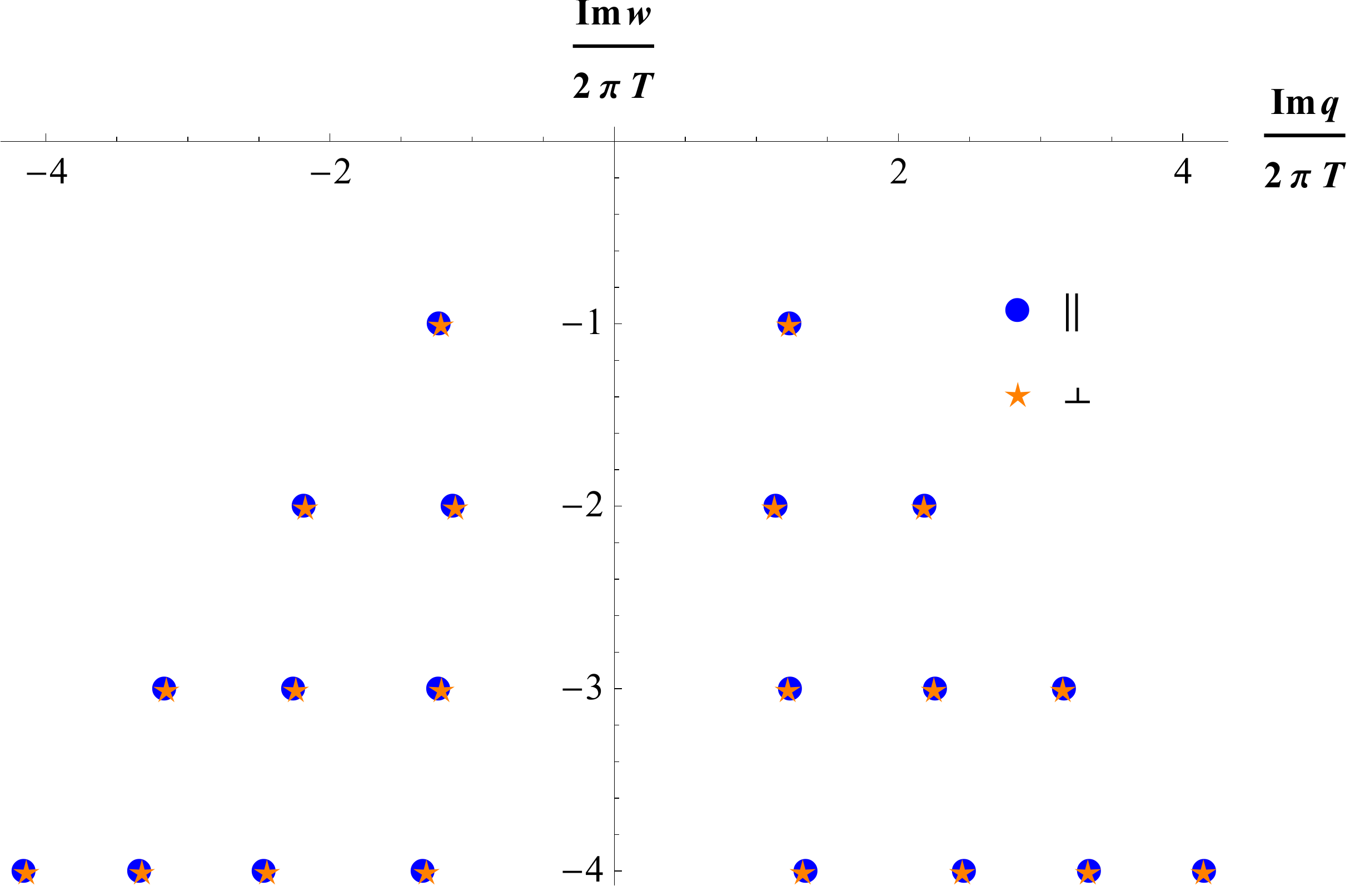}
\end{tabular}
\caption{ a) Locations of the pole skipping points for the scalar field obtained by means of numerical calculations with the dimensionless parameter $b=0.1$, b) Plot showing the pole skipping points obtained numerical calculations for the axion field with $b=0.15$.}
\label{1}
\end{figure}
In Figure-{\bf 1a} we manage to plot the first four pole skipping points for the scalar field in the complex $(w-q)$ plane for both the cases with perturbation propagation parallel (denoted by the blue dots) and perpendicular (denoted by the orange star) to the direction of anisotropy. We see from the plot that the results for the parallel and the perpendicular case differ by very small amount. Also note that, the value of $w$ at the special points at different order does not receive any correction due to non zero anisotropy but only the value of $q$ gets a finite correction.
\begin{figure}[h]
\centering
\begin{tabular}{ccc}
\includegraphics[width=.45\textwidth]{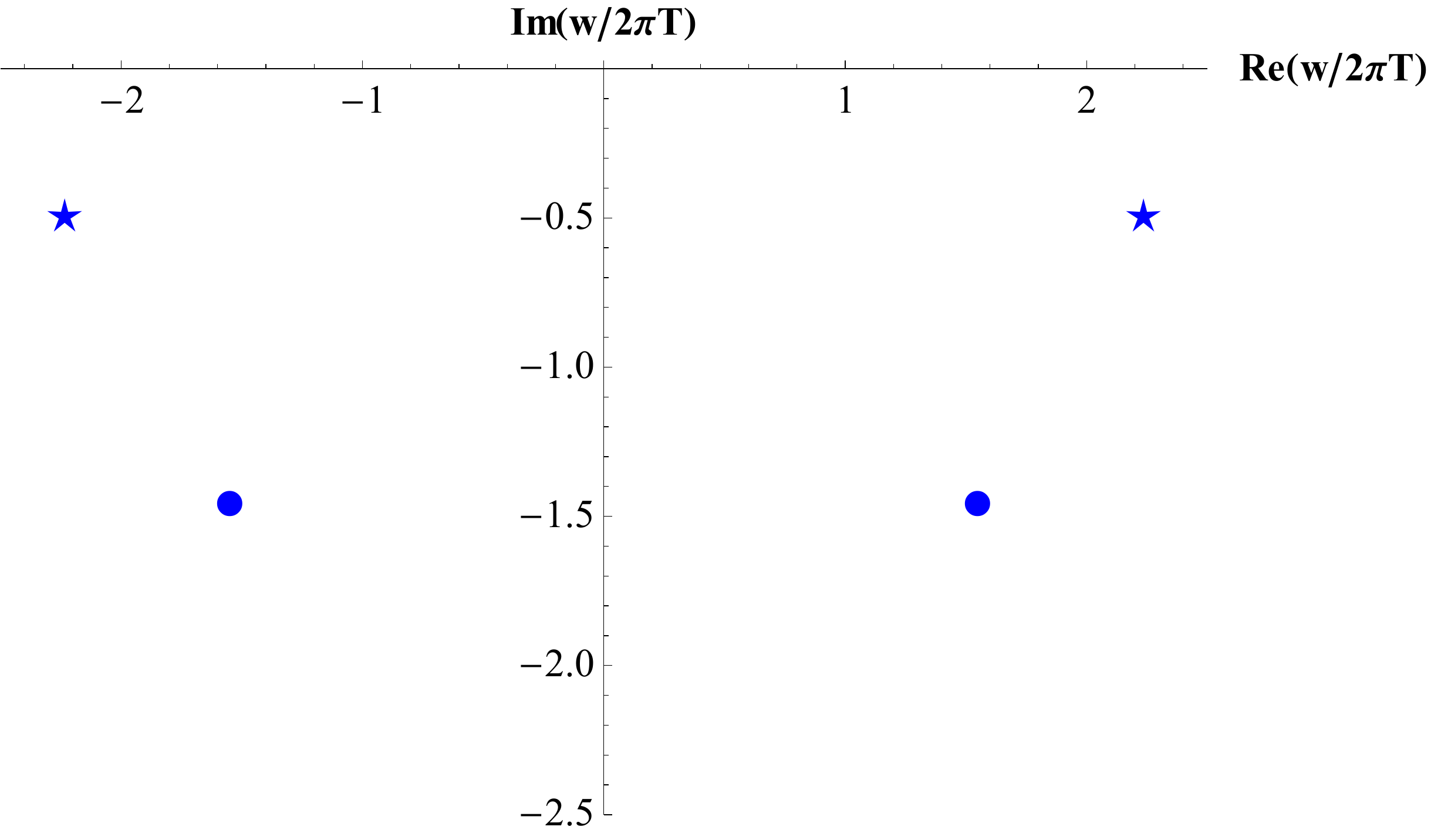}&&
\includegraphics[width=.45\textwidth]{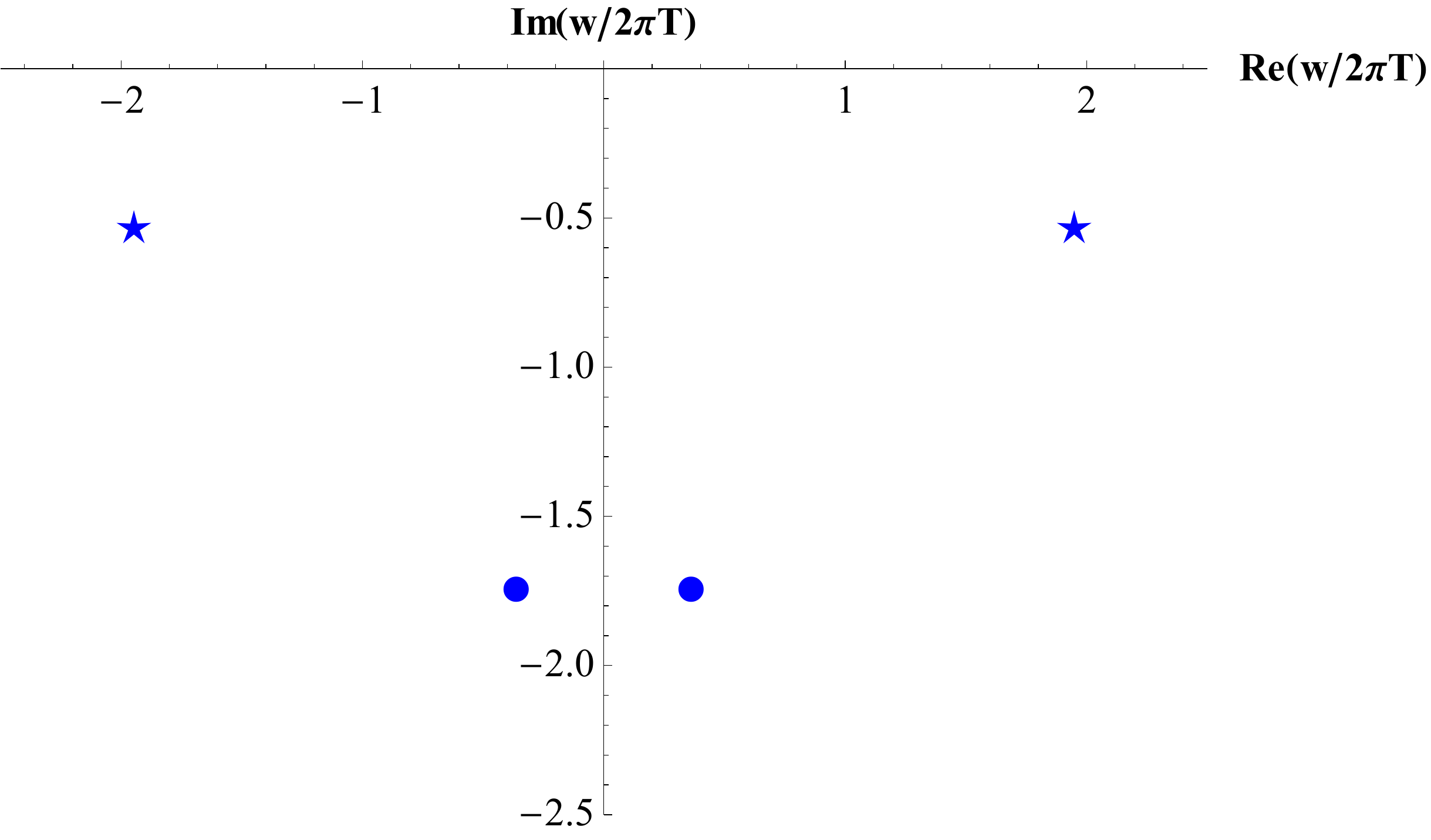}\\
\includegraphics[width=.45\textwidth]{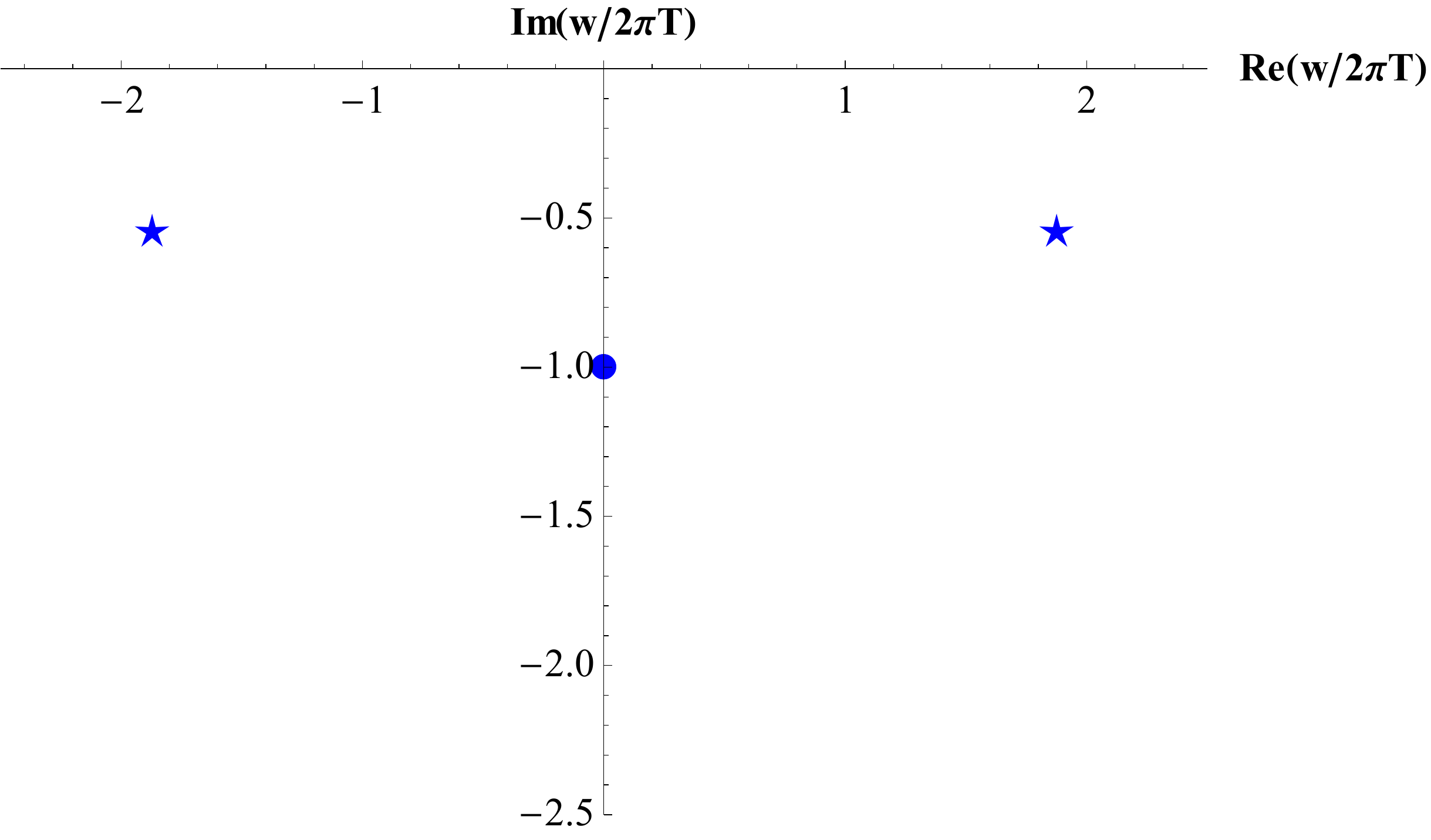}&&
\includegraphics[width=.45\textwidth]{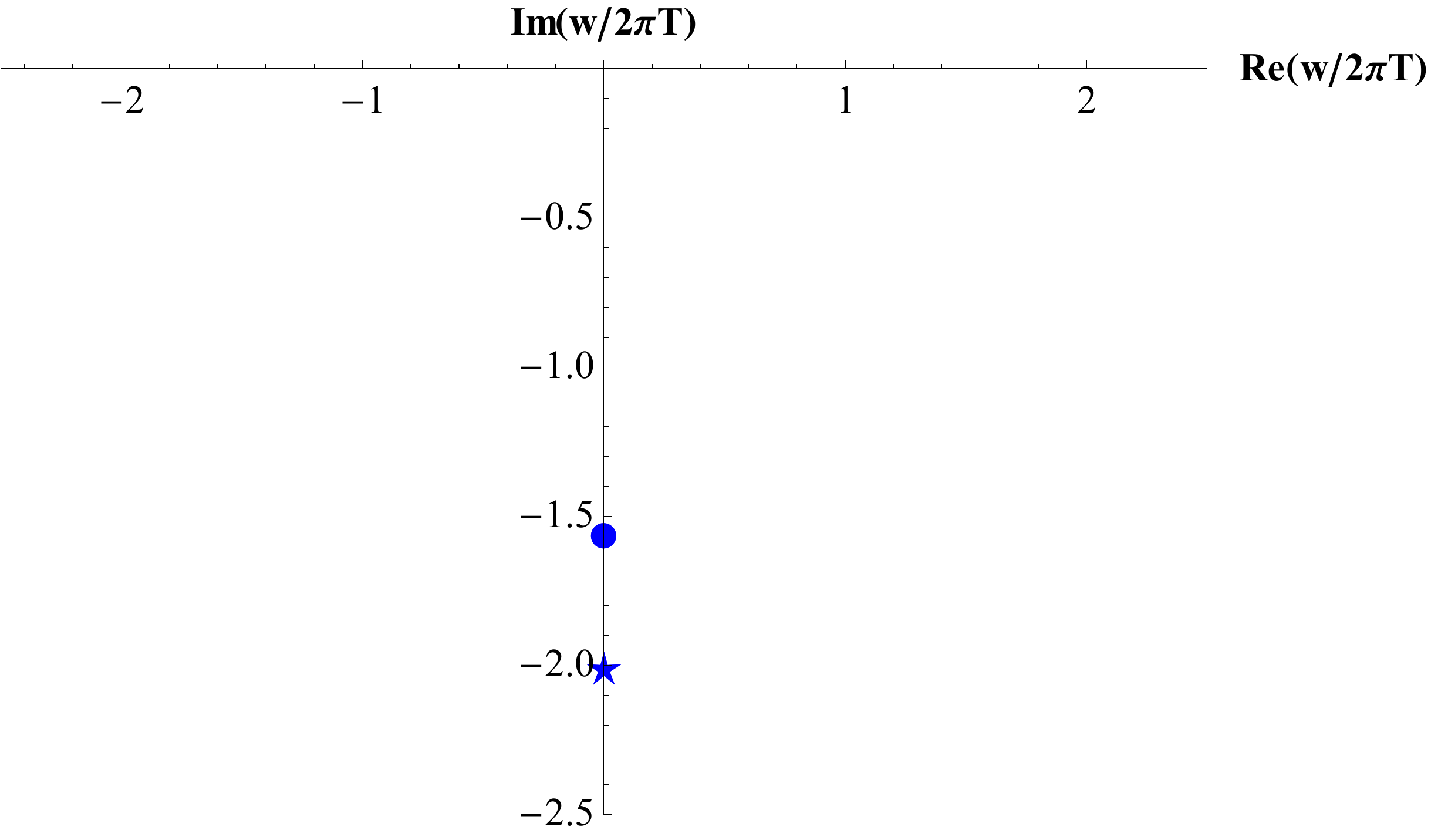}
\end{tabular}
\caption{Plot showing the movement of the poles of retarded Green's function as the value of the momentum is increased from zero to finite values. }
\label{1}
\end{figure}

Next we compute the pole of the retarded Green's function corresponding to some scalar operator
which is dual to the bulk scalar field considered above. Using the numerical methods as described
in \cite{Horowitz:1999jd, Kovtun:2005ev} (we discuss this numerical method in some details in section-4), the pole of the Green's
function can be obtained. In Figure-{\bf 2} we have shown the locations of the poles occurring at first and second order as denoted by dot and star symbol respectively. Different plots in the same figure corresponds to four different values of the dimensionless momentum, namely the top left plot corresponds to the dimensionless momentum $Q=\frac{q}{2\pi T}=0$, the top right with $Q=1.1 i$, the bottom left with $Q=1.225 i$ and the bottom right with $Q=1.126 i$. The last two values of $Q$ corresponds to the pole skipping points at first and second order respectively. Hence at $Q=1.225 i$ and $Q=1.126 i$ the pole has to appear at $W=-1 i$ and $W=-2 i$ respectively for $W=\frac{w}{2\pi T}$ which is clearly evident from Figure-{\bf 2}.
\subsection{Axion field fluctuation}
The equation of motion for the axion field fluctuation $\chi$ is almost similar to that of the scalar field discussed above. Also after performing the near horizon analysis, the components of the matrix $C$ in (\ref{Cmatrix}) turns out to be very similar to those obtained for the scalar field fluctuation. In fact the leading ordered terms appearing in $C$ are exactly the same. So we will not write them down again but only mention the final results for the pole skipping points at different order. Again the pole skipping points at first and second order can be solved exactly as given below for the two different cases with the perturbation being parallel and perpendicular to the direction of anisotropy.
\begin{itemize}
\item {\it Parallel case:}
\end{itemize}
\begin{equation}\begin{split}
&w_{1}=-i2\pi T,~~~~q_{1}^2=-6r_{H}^2-\frac{a^2}{8}\biggl(2-18\log{2}\biggr),\\&
w_{2}=-i4\pi T,~~~~~q_{2}^2=\biggl(-12\pm4\sqrt{3}\biggr)r_{H}^2+\frac{a^2}{2}\biggl(2\mp\sqrt{3}-3\left(3\mp\sqrt{3}\right)\log 2\biggr)
\end{split}
\end{equation}
\begin{itemize}
\item {\it Perpendicular case:}
\end{itemize}
\begin{equation}\begin{split}
&w_{1}=-i2\pi T,~~~~q_{1}^2=-6r_{H}^2+\frac{a^2}{4}\biggl(1-3\log{2}\biggr),\\&
w_{2}=-i4\pi T,~~~~~q_{2}^2=\biggl(-12\pm4\sqrt{3}\biggr)r_{H}^2+\frac{a^2}{2}\biggl(1-3\left(3\mp\sqrt{3}\right)\log 2\biggr)
\end{split}
\end{equation}
The higher order pole skipping points are solve them by numerical methods. We have shown the locations of those points in Figure-{\bf 1b} with the blur colored dots and the orange stars indicating the parallel and perpendicular cases respectively.
\subsection{Metric Perturbation}
Let us now discuss the metric field perturbations with two different modes of perturbations, the vector and scalar modes which corresponds respectively to shear and sound channel. The non zero components are discussed in section-{\bf 3} for both parallel and perpendicular case. However in order to make the calculations simpler we will work in a particular gauge where all the metric perturbations $h_{r\mu}$ for all $\mu$ is set to zero. Einstein's equation for the two modes of perturbations can be cast into a closed form in terms of a single equation involving the gauge invariant variable. The constructions of the gauge invariant variables are discussed in Appendix-{\bf A}.
\subsubsection{Shear channel}
In this section we consider the vector modes of metric perturbation. Similar to the scalar and axion field, we will consider the following two separate cases,
\begin{itemize}
\item{\it Parallel case}
\end{itemize}
In the particularly chosen gauge we have only two non zero components for the field field perturbation in this case. They are defined in the fourier space as,
\begin{equation}
h_{vx_{2}}=e^{-iwv+iqx_{1}}g_{vv}H_{vx_{2}}(r),~~~~~h_{x_{1}x_{2}}=e^{iwv+iqx_{1}}g_{11}H_{x_{1}x_{2}}(r),
\end{equation}
with $H_{vx_{2}}=h_{vv}/g_{vv}$ and $H_{x_{1}x_{2}}=h_{x_{1}x_{2}}/g_{11}$. The two independent Einstein's equations can be written in the following form,
\begin{equation}\label{vecvec}
\begin{split}
&H_{x_{1}x_{2}}^{\prime}=\biggl(F_{x_{1}x_{2}}+a^2\tilde{F}_{x_{1}x_{2}}\biggr)H_{vx_{2}}^{\prime}+\biggl(G_{x_{1}x_{2}}+a^2\tilde{G}_{x_{1}x_{2}}\biggr)
H_{vx_{2}}+\biggl(J_{x_{1}x_{2}}+a^2\tilde{J}_{x_{1}x_{2}}\biggr)H_{x_{1}x_{2}},\\&
H_{vx_{2}}^{\prime\prime}=\biggl(F_{vx_{2}}+a^2\tilde{F}_{vx_{2}}\biggr)H_{vx_{2}}^{\prime}+\biggl(G_{vx_{2}}+a^2\tilde{G}_{vx_{2}}\biggr)
H_{vx_{2}}+\biggl(J_{vx_{2}}+a^2\tilde{J}_{vx_{2}}\biggr)H_{x_{1}x_{2}},
\end{split}
\end{equation}
where all the coefficients $F_{x_{1}x_{2}}, G_{x_{1}x_{2}}, J_{x_{1}x_{2}}..$ are functions of $w, q$ and $r_{H}$. We will not write their exact form as the expressions are too long.
Using the gauge invariant variable $Z_{v\parallel}$ as given in (\ref{GIvec}), the above two equations can be clubbed into a single equation involving $Z_{v\parallel}$ which is given as,
\begin{equation}\label{veceqn}
Z_{v\parallel}^{\prime\prime}-\left(\mathcal{N}+a^2\tilde{\mathcal{N}}\right)Z_{v\parallel}^{\prime}
-\left(\mathcal{P}+a^2\tilde{\mathcal{P}}\right)Z_{v\parallel}=0,
\end{equation}
where the coefficients $\mathcal{N}$, $\tilde{\mathcal{N}}$, $\mathcal{P}$ and $\tilde{\mathcal{P}}$ in the above equation are given in Appendix-{\bf C}. We consider a near horizon expansion for $Z_{v\parallel}$ as given in (\ref{pex}) and substitute it
into equation (\ref{veceqn}) to get the first few components of the matrix $C$ in (\ref{Cmatrix}),
\begin{equation}
\begin{split}
&c_{00}=\frac{1}{192wr_{H}^5}\biggl[-48r_{H}^2\biggl(wr_{H}(8r_{H}-iw)+q^2(4ir_{H}+w)\biggr)
\\&+a^2\biggl\{wr_{H}\biggl(24r_{H}-iw(7+\log{(32)})\biggr)
+q^2\biggl(3w(1+\log{(64)})+4ir_{H}(5+\log{(8192)})\biggr)\biggr\}\biggr]\\
&c_{01}=\frac{1}{96r_{H}^3}\biggl[96r_{H}^3-48iwr_{H}^2+a^2\biggl(iw\left(-2+5\log(2)\right)\biggr)\biggr]\\
&c_{10}=\frac{1}{384w^3r_{H}^6}\biggl[-48ir_{H}^2\biggl(32q^4r_{H}
+w^3r_{H}(24ir_{H}+w)+iwq^2\left(-64r_{H}^2+36iwr_{H}+3w^2\right)\biggr)
\\&+a^2\biggl\{2wq^2\biggl(9w^2(2+\log{(8)})-64r_{H}^2(5+\log(16))+18iwr_{H}(14+\log{(8192)})\biggr)\\&+
w^3r_{H}\biggl(-288r_{H}+iw(16+\log{(32)})\biggr)\biggr\}\biggr]\\
&c_{11}=\frac{1}{192w^2r_{H}^5}\biggl[48r_{H}^2\biggl(-14w^2r_{H}^2
+q^2(16r_{H}^2-4iwr_{H}-w^2)\biggr)
+a^2\biggl\{78w^2r_{H}^2\\&+q^2\biggl(3w^2(1+\log{(64)})
-16r_{H}^2(7+\log{(256)})+4iwr_{H}(5+\log{(8192)})\biggr)\biggr\}\biggr]\\
&c_{12}=\frac{1}{48r_{H}^3}\biggl[192r_{H}^3-48iwr_{H}^2+a^2\biggl(iw\left(-2+5\log(2)\right)\biggr)\biggr]
\end{split}
\end{equation}
Solving (\ref{matrixeqn}), the first two pole skipping points are obtained as,
\begin{equation}
\begin{split}
& w_1=-i 2\pi T ~~~~q_{1}^{2}=6r_{H}^2+\frac{a^2}{4}\biggl(1+\log(512)\biggr),\\
& w_2=-i 4\pi T,~~~~q^2_2=\pm 4\sqrt{6}r_{H}^{2}+a^2\biggl(1\pm\sqrt{\frac{3}{2}}\log(8)\biggr),
\end{split}
\end{equation}
Notice that here the pole skipping points corresponds to real momentum along with the imaginary solutions. These real momentum puts nontrivial constrains to the transverse momentum dispersion relation which will be discussed in the next subsection.
\begin{itemize}
\item{\it Perpendicular case}
\end{itemize}
In this case the non zero components of the field perturbations are,
\begin{equation}
h_{vx_{3}}=e^{-iwv+iqx_{2}}g_{vv}H_{vx_{3}}(r),~~~~~h_{x_{2}x_{3}}=e^{iwv+iqx_{2}}g_{22}H_{x_{2}x_{3}}(r),
\end{equation}
Similar to the parallel case discussed above one gets two independent equations which can be clubbed together using the gauge invariant variable as given in (\ref{GIvec1}). The equation of motion involving the gauge invariant variable $Z_{v\perp}$ is given as,
\begin{equation}\label{veceqn2}
Z_{v\perp}^{\prime\prime}-\left(\mathcal{N}+a^2\tilde{\tilde{\mathcal{N}}}\right)Z_{v\perp}^{\prime}
-\left(\mathcal{P}+a^2\tilde{\tilde{\mathcal{P}}}\right)Z_{v\perp}=0,
\end{equation}
The power series ansatz near the horizon gives the following few components of the matrix $C$ in (\ref{Cmatrix}),
\begin{equation}
\begin{split}
&c_{00}=\frac{1}{192wr_{H}^5}\biggl[-48r_{H}^2\biggl(wr_{H}(8r_{H}-iw)+q^2(4ir_{H}+w)\biggr)
\\&+a^2\biggl\{wr_{H}\biggl(24r_{H}-5iw(-1+\log{(2)})\biggr)
+q^2\biggl(w(3+\log{(64)})+4ir_{H}(5+\log{(2)})\biggr)\biggr\}\biggr]\\
&c_{01}=\frac{1}{96r_{H}^3}\biggl[96r_{H}^3-48iwr_{H}^2+a^2\biggl(iw\left(-2+5\log(2)\right)\biggr)\biggr]\\
&c_{10}=\frac{1}{384w^3r_{H}^6}\biggl[-48ir_{H}^2\biggl(32q^4r_{H}
+w^3r_{H}(24ir_{H}+w)+iwq^2\left(-64r_{H}^2+36iwr_{H}+3w^2\right)\biggr)
\\&+a^2\biggl\{-2wq^2\biggl(3w^2(2+\log{(8)})+64r_{H}^2(-5+\log(4))+6iwr_{H}(26+\log{(8)})\biggr)\\&+
w^3r_{H}\biggl(-96r_{H}+5iw(-4+\log{(2)})\biggr)\biggr\}\biggr]\\
&c_{11}=\frac{1}{192w^2r_{H}^5}\biggl[48r_{H}^2\biggl(-14w^2r_{H}^2
+q^2(16r_{H}^2-4iwr_{H}-w^2)\biggr)
+a^2\biggl\{6w^2r_{H}(5r_{H}+2iw)\\&+q^2\biggl(w^2(3+\log{(64)})
+16r_{H}^2(-7+\log{(16)})+4iwr_{H}(5+\log{(2)})\biggr)\biggr\}\biggr]\\
&c_{12}=\frac{1}{48r_{H}^3}\biggl[192r_{H}^3-48iwr_{H}^2+a^2\biggl(iw\left(-2+5\log(2)\right)\biggr)\biggr]
\end{split}
\end{equation}
yielding the following analytic results for the first two pole skipping points,
\begin{equation}
\begin{split}
& w_1=-i 2\pi T ~~~~q_{1}^{2}=6r_{H}^2+\frac{a^2}{8}\biggl(-2+\log(64)\biggr),\\
& w_2=-i 4\pi T,~~~~q^2_2=\pm 4\sqrt{6}r_{H}^{2}\pm \frac{a^2}{4\sqrt{6}}\biggl(-6+\log(4096)\biggr),
\end{split}
\end{equation}
\subsubsection{Transverse momentum diffusion in shear channel}
In this section we wish to evaluate explicitly the location of the diffusion poles in the complex $(w-q)$ plane which according to the phenomenon of pole skipping is constrained to pass through the pole skipping points as obtained in the previous subsection. In particular, we are interested in the dispersion relation that arises from the pole of the retarded Green's function associated to the transverse momentum density. The non zero components for the perturbed fields in this case again will be of vector type which in $(t,u=\frac{r_{H}^2}{r^2},\vec{x})$ coordinates are $h_{tx_{2}}$ and $h_{x_{1}x_{2}}$ (again we consider a particular gauge such that $h_{\mu u}=0$, for all $\mu$) with the perturbation propagating along the anisotropic direction $x_{1}$ as $h_{tx_{1}/x_{1}x_{2}}(t,u,x_{1})=e^{-iwt+iqx_{1}}h_{tx_{1}/x_{1}x_{2}}(u)$. The Einstein's equation in this case can be put in a closed form in terms of the gauge invariant variable, $\mathcal{Z}_{v}=w h_{tx_{1}}+q h_{x_{1}x_{2}}$ as,
\begin{equation}\label{difusion}
\mathcal{Z}^{\prime\prime}_{v}-\biggl(\mathcal{R}+b^2\tilde{\mathcal{R}}\biggr)\mathcal{Z}^{\prime}_{v}-\biggl(\mathcal{S}+b^2\tilde{\mathcal{S}}\biggr)\mathcal{Z}_{v}=0,
\end{equation}
with the coefficients $\mathcal{R}, \mathcal{S}, \tilde{\mathcal{R}}, \tilde{\mathcal{S}}$ as given in (\ref{difucoff}). Here as before, we define dimensionless frequency $W$ and momentum $Q$ as $W=(w/2\pi T)$ and $Q=(q/2\pi T)$.
Now, to determine the dispersion relation we follow the numerical approach as given in \cite{Horowitz:1999jd, Kovtun:2005ev}. First, the behavior near the horizon is determined by inserting the ansatz $(1-u)^{\alpha}$ into equation (\ref{difusion}). Two possible solutions for $\alpha$ is obtained as $\alpha=\pm iW/2$ in which the solution with the negative sign is chosen to impose the incoming wave boundary condition at the horizon. Then the final solution can be written as the following power series,
\begin{equation}
\mathcal{Z}_{v}=(1-u)^{-\frac{iW}{2}}\sum_{n=0}^{\infty}c_{n}(W,Q)\biggl(1-u\biggr)^n.
\end{equation}
Imposing the following Dirichlet boundary condition at the boundary $(u=0)$ one determines the quasinormal modes,
\begin{equation}
\mathcal{Z}_{v}(0)=\sum_{n=0}^{\infty}c_{n}(W,Q)=0.
\end{equation}
In Figure-{\bf 3}, we have shown the quasinormal modes for the exact dispersion relation (blue dots) as obtained by the above numerical method. Notice that the blue dots which are the poles of the correlation function, passes through the pole skipping points (only the first three points are shown in the graph with locations given by the points $(-1i,1.2284)$, $(-2i, 1.5710)$, $(-3i, 1.8041)$ where the first value represents the complex frequency and the second one is the real momentum) which are represented by points where the horizontal and vertical black dashed lines intersects each other. The pole structure of the retarded Green's function for transverse momentum density at very small momentum and frequency $Q, W<<1$ (hydrodynamic approximation) gives the following dispersion relation,
\begin{equation}
W=-i D Q^2+...
\end{equation}
$D$ being the diffusion constant. However, the above equation is not appropriate at large energy scale, $W\sim T$. The above discussion shows that the behavior of the dispersion relation at large energy can be predicted from a simple near horizon analysis due to the pole skipping phenomenon.
\begin{figure}[h]
\centering
\begin{tabular}{c}
\includegraphics[width=.55\textwidth]{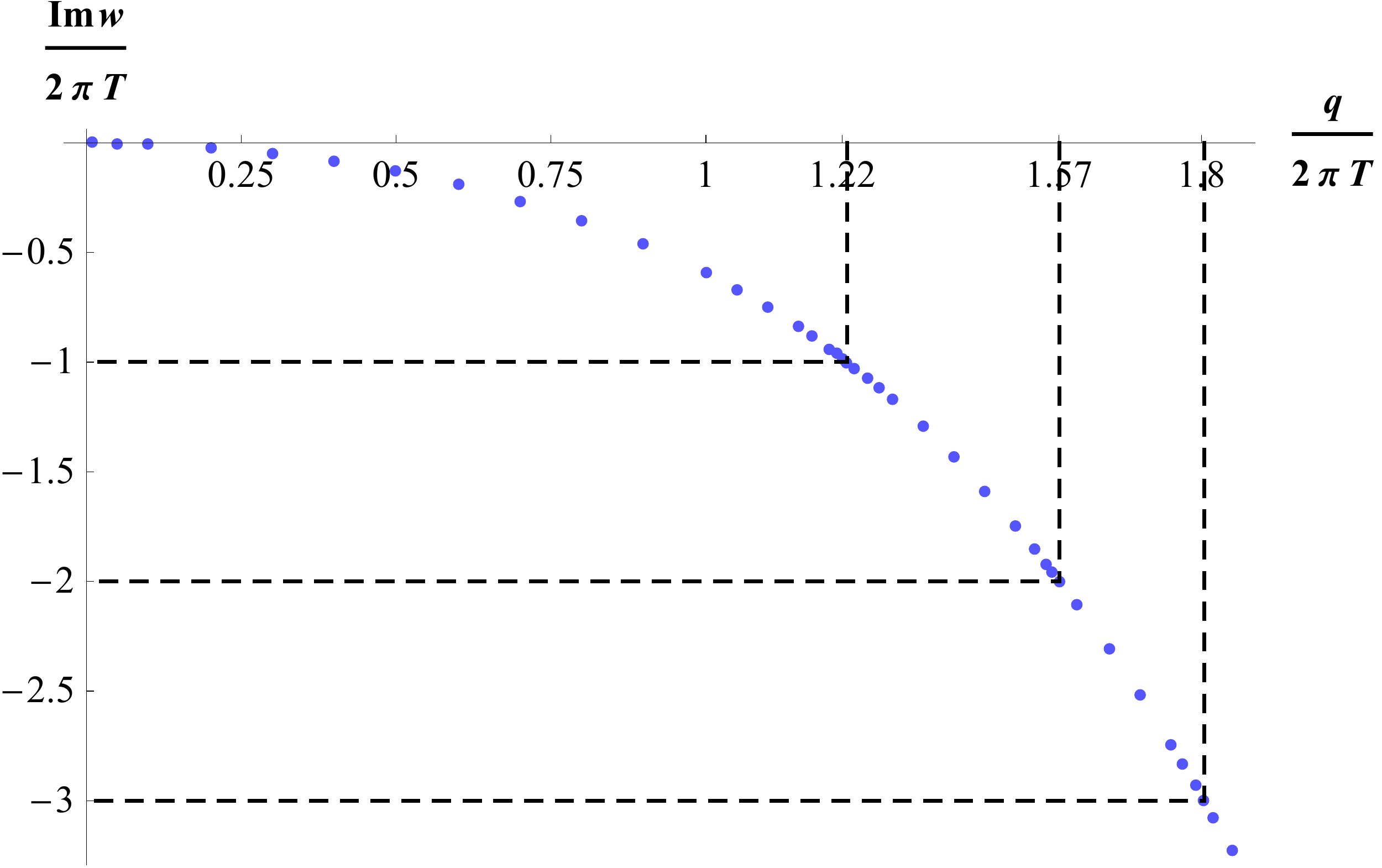}
\end{tabular}
\caption{Plot showing the dispersion relation for momentum diffusion which passes through the first three successive pole skipping points. The calculation for the dispersion relation is done using numerical methods with $b=.5$}
\label{1}
\end{figure}
\subsubsection{Sound channel}
The non zero components of the metric perturbations for the scalar modes are already given in equation (\ref{parallel}) and (\ref{perp}) respectively for the perturbation propagating along or normal to the direction of anisotropy. Here, to make the discussion simpler we will consider a particular gauge so that all the metric fluctuations of the form $h_{r\mu}$ for all $\mu$ will be set to zero.
\begin{itemize}
\item{\it Parallel case}
\end{itemize}
Considering the scalar modes of metric perturbation the full set of fluctuations are given as,
\begin{equation}\label{fluc}
\begin{split}
& ~~h_{vv}=e^{-iwv+iqx_1}g_{vv}H_{vv}(r),~~~~~h_{v x_{1}}=e^{-iwv+iqx_1}g_{11}H_{vx_{1}}(r),\\& h_{x_{1}x_{1}}=e^{-iwv+iqx_1}g_{11}H_{x_{1}x_{1}}(r),
~~h_{x_{2}x_{2}}=e^{-iwv+iqx_1}g_{22}H_{x_{1}x_{1}}(r),
\end{split}
\end{equation}
where in the above we define $H_{vv}=h_{vv}/g_{vv}, H_{vx_{1}}=h_{vx_{1}}/g_{11}, H_{x_{1}x_{1}}=h_{x_{1}x_{1}}/g_{11}, H_{x_{2}x_{2}}=h_{x_{2}x_{2}}/g_{22}$.
Now, using the above form of different fluctuations into the linearized equations (\ref{lineareqn}), we obtain the following four linearly independent coupled differential equations for the metric perturbations as,
\begin{equation}\label{eq1}
\begin{split}
H_{mn}^{\prime}&=(A_{mn}+a^2\tilde{A}_{mn})H_{vv}^{\prime}+(B_{mn}+a^2\tilde{B}_{mn})H_{vv}+(C_{mn}+a^2\tilde{C}_{mn})H_{vx_{1}}
\\&+(D_{mn}+a^2\tilde{D}_{mn})H_{x_{1}x_{1}}+(E_{mn}+a^2\tilde{E}_{mn})H_{x_{2}x_{2}}
\\ H_{vv}^{\prime\prime}&=(A_{vv}+a^2\tilde{A}_{vv})H_{vv}^{\prime}+(B_{vv}+a^2\tilde{B}_{vv})H_{vv}+(C_{vv}+a^2\tilde{C}_{vv})H_{vx_{1}}
\\&+(D_{vv}+a^2\tilde{D}_{vv})H_{x_{1}x_{1}}+(E_{vv}+a^2\tilde{E}_{vv})H_{x_{2}x_{2}},
\end{split}
\end{equation}
where, $H_{mn}=\left\{H_{vx_{1}},H_{x_{1}x_{1}},H_{x_{2}x_{2}}\right\}$ and all the coefficients in the above equation namely, $A_{mn}, B_{mn},C_{mn}$....etc are too lengthy to write in the paper but are functions of $(r,w,q)$. We can see that all the equations written above in (\ref{eq1}) are coupled which makes it difficult to solve them.
However constructing gauge invariant variables \cite{Benincasa:2005qc, Waeber:2015oka, Kovtun:2005ev, Parnachev:2005hh, Cai:2016sur, Benincasa:2005iv} by combining the field fluctuations one can reduce the coupled equations into a single equation involving the gauge invariant variables. The details of the construction of the gauge invariant variables for scalar and vector modes of metric perturbations $\{(Z_{s}),(Z_{\mathrm{v}})\}$ for the anisotropic gravitational background are given in Appendix-{\bf A}. The equation of motions involving the gauge invariant variables for the metric perturbations (scalar modes) is given as \footnote{See \cite{Czajka:2018bod, Czajka:2018egm, Sil:2016jmc} for the detailed procedure to obtain the equation of motion involving the gauge invariant variable.},
\begin{equation}\label{GIqnn}
\begin{split}
&Z^{\prime\prime}_{s\parallel}-\left(\mathcal{M}+a^2\tilde{\mathcal{M}}\right)Z_{s\parallel}^{\prime}-\left(\mathcal{L}+a^2\tilde{\mathcal{L}}
\right)Z_{s\parallel}=0,
\end{split}
\end{equation}
We write down the exact expression for $\mathcal{M}$ and $\mathcal{L}$ in (\ref{coeff}) and (\ref{coeff2}) respectively, while the results for $\tilde{\mathcal{M}}$ and $\tilde{\mathcal{L}}$ are given in (\ref{coeff1}) and (\ref{coeff3}). Here we are interested to find the pole skipping points in the upper half complex plane and for this we closely follow the analysis done in \cite{Blake:2019otz, Wu:2019esr}.

Following the results as obtained in section-{\bf 3}, the location of the special point in the absence of the anisotropy $(a=0)$, is given from the near horizon analysis as,
\begin{equation}\label{wqnoaniso}
w=2i r_{H},~~~~~ q=\sqrt{\frac{3}{2}}w.
\end{equation}
In order to proceed with the near horizon analysis of equation (\ref{GIqnn}), we must check its singularity structure near $r=r_{H}$ which changes at the special location as given in the above equation. In particular at $q=\sqrt{3/2}w$, in the near horizon limit $\mathcal{M}$ and $\mathcal{L}$ in the above equation is dominated by terms proportional to $1/(r-r_{H})$ and $1/(r-r_{H})^2$ respectively. In presence of the anisotropy which is considered in a perturbative approximation ($a<<T$ or $b<<1$), equation (\ref{GIqnn}) must abide by the above mentioned regularity condition at the special point. In other words, any term that appears in the near horizon expansion of $\left(\mathcal{M}+a^2\tilde{\mathcal{M}}\right)$ and $\left(\mathcal{L}+a^2\tilde{\mathcal{L}}
\right)$ which is proportional to $\left(1/(r-r_{H})^{p}\right)$ with $p\ge 2$ and $p\ge 3$ respectively must be equated to zero \cite{Wu:2019esr}.

Turning on the anisotropy, we expect the special point to get shifted from the value mention in the above equation (\ref{wqnoaniso}). Let us assume that the coordinate of the shifted point in the complex $(w-q)$ plane is given as,
\begin{equation}\label{wqnew}
q=\sqrt{\frac{3}{2}}w+a^{2}q_{1},~~~~~~~w=2i r_{H}+a^{2}w_{1},
\end{equation}
where we required to determine $w_{1}$ and $q_{1}$.
Inserting the above choice for $w,q$ into (\ref{ml}) we obtain the following near horizon expansion for the coefficient of $Z_{s\parallel}^{\prime}$ in (\ref{GIqnn}) as,
\begin{equation}
\left(\mathcal{M}+a^2\tilde{\mathcal{M}}\right)\sim \frac{a^2\left(1+i 8\sqrt{6}q_{1}r_{H}+4\log(2)\right)}{24  r_H\left(r-r_H\right)^2}+\mathcal{O}\left((r-r_{H})^{-1}\right).
\end{equation}
Equating the above to zero, we obtain the following result for $q_{1}$ as,
\begin{equation}\label{q1}
q_{1}=\frac{i\left(1+4\log(2)\right)}{8 \sqrt{6} r_H}.
\end{equation}
On the other hand, the near horizon expansion for the coefficient of $Z_{s\parallel}$ in (\ref{GIqnn}) again with the shifted $w$ and $q$ is already dominated by term $\mathcal{O}\left((r-r_{H})^{-2}\right)$.
Using the obtained result for $q_{1}$ and keeping only the most dominating terms for the near horizon expansion of $\left(\mathcal{M}+a^2\tilde{\mathcal{M}}\right)$ and $\left(\mathcal{L}+a^2\tilde{\mathcal{L}}
\right)$ one gets,
\begin{equation}\label{eqnn}
\begin{split}
\left(\mathcal{M}+a^2\tilde{\mathcal{M}}\right)&=\frac{a^2\biggl(-28+23\log{(2)}+24i w_{1}r_{H}\biggr)}{48r_{H}^2(r-r_{H})}+\mathcal{O}\biggl((r-r_{H})^{0}\biggr)\\
\left(\mathcal{L}+a^2\tilde{\mathcal{L}}
\right)&=\frac{a^2\biggl(2-5\log{(2)}-24i w_{1}r_{H}\biggr)}{48r_{H}^2(r-r_{H})^2}+\mathcal{O}\left((r-r_{H})^{-1}\right),
\end{split}
\end{equation}
Now, we consider the following power series ansatz for $Z_{s\parallel}$,
\begin{equation}\label{pansatz}
Z_{s\parallel}=(r-r_{H})^{\lambda}\sum_{n=0}Z_{sn}(r-r_{H})^{n}.
\end{equation}
Inserting the above in (\ref{GIqnn}) with the coefficients of $Z_{s\parallel}^{\prime}$, $Z_{s\parallel}$ as given in (\ref{eqnn}) and solving the indicial equation for $\lambda$, one gets the following two solutions,
\begin{equation}\label{w1}
\begin{split}
\lambda_{1}&=\frac{a^2\biggl(-2+5\log{(2)}+24iw_{1}r_{H}\biggr)}{48r_{H}^2},\\
\lambda_{2}&=1+\frac{a^2\biggl(-13+9\log{(2)}\biggr)}{24r_{H}^2}
\end{split}
\end{equation}
Solving for $w_{1}$ such that $\lambda_{1}=0$, we get the final result for $w$ and $q$ from (\ref{wqnew}) as,
\begin{equation}\begin{split}
w=2 i\pi T,~~~~~
q=i\sqrt{6}\pi T\left[1+b^2\left(\frac{1+4\log(2)}{48\pi^2}\right)\right]
\end{split}
\end{equation}
\begin{itemize}
\item{\it Perpendicular case}
\end{itemize}
Let us now consider the perturbation along the $x_2$ direction.
The non zero components of the perturbations in this case are given as,
\begin{equation}\label{fluc1}
\begin{split}
& ~~h_{vv}=e^{-iwv+iqx_2}g_{vv}H_{vv}(r),~~~~~h_{v x_{2}}=e^{-iwv+iqx_2}g_{22}H_{vx_{2}}(r),\\& h_{x_{1}x_{1}}=e^{-iwv+iqx_2}g_{11}H_{x_{1}x_{1}}(r),
~~h_{x_{2}x_{2}}=e^{-iwv+iqx_2}g_{22}H_{x_{2}x_{2}}(r),~~~~h_{x_3x_3}=\left(g_{22}/g_{11}\right)h_{x_1x_1},
\end{split}
\end{equation}
Again with the above perturbations we will get four linearly independent equations similar to (\ref{eq1}) which can be put in a closed form using the gauge invariant variable obtained in (\ref{gaugeinvzs2}).
The equation in terms of the gauge invariant variable $Z_{s\perp}$ is given as,
\begin{equation}\label{GIqnn2}
\begin{split}
Z^{\prime\prime}_{s\perp}-\left(\mathcal{M}+a^2\tilde{\tilde{\mathcal{M}}}\right)Z_{s\perp}^{\prime}-\left(\mathcal{L}+a^2\tilde{\tilde{\mathcal{L}}}
\right)Z_{s\perp}=0.
\end{split}
\end{equation}
The expressions of $\tilde{\tilde{\mathcal{M}}}$ and $\tilde{\tilde{\mathcal{L}}}$ are given in (\ref{coeff4}) and (\ref{coeff5}) respectively. The analysis towards the final results for $w$ and $q$ are exactly similar to what we have done in the previously corresponding to the perturbation that is parallel to the direction of anisotropy. In particular in this case also we found the coefficient of $Z_{s\perp}^{\prime}$ to behave near the horizon as $\sim 1/(r-r_{H})^2$ which must vanish in order for the perturbative analysis to be consistent. Considering the same ansatz for $w$ and $q$ as given in (\ref{wqnew}), the near horizon behavior of $\left(\mathcal{M}+a^2\tilde{\tilde{\mathcal{M}}}\right)$ is given as,
\begin{equation}
\left(\mathcal{M}+a^2\tilde{\tilde{\mathcal{M}}}\right)\sim \frac{a^2\biggl(1+i 8\sqrt{6}q_{1}r_{H}-2\log(2)\biggr)}{24  r_H\left(r-r_H\right)^2}+\mathcal{O}\left((r-r_{H})^{-1}\right).
\end{equation}
The above vanishes exactly for the value of $q_{1}$ given as,
\begin{equation}\label{q2}
q_{1}=-i\frac{\biggl(-1+2\log(2)\biggr)}{8\sqrt{6} r_H}.
\end{equation}
Again considering the power series ansatz for $Z_{s\perp}$ similar to the one as given in (\ref{pansatz}) and solving the indicial equation the exponent $\lambda$ can be solved as,
\begin{equation}
\begin{split}
\lambda_{1}&=\frac{a^2\biggl(-2+5\log{(2)}+24iw_{1}r_{H}\biggr)}{48r_{H}^2},\\
\lambda_{2}&=1+\frac{a^2\biggl(-2+\log{2}\biggr)}{8r_{H}^2},
\end{split}
\end{equation}
such that, the value of $w_{1}$ remains the same as in (\ref{w1}), with the final results given as,
\begin{equation}\label{gtggt}
w=2i\pi T,~~~~~q=i\sqrt{6}\pi T\left[1-b^2\left(\frac{-1+2\log(2)}{48\pi^2}\right)\right],
\end{equation}
The results for the pole skipping points in the sound channel as obtained in this section using the gauge invariant approach is exactly matches with the results in section-{\bf 3} for both parallel and perpendicular case.
\section{Conclusion}
In the current manuscript we have done a detailed analysis of the very recently observed phenomenon called "Pole skipping" in a strongly coupled plasma with anisotropy along a spatial direction from the near horizon analysis of the equation of motions for different bulk field perturbations. We have also shown that this phenomenon helps us determine the parameters of chaos for the same anisotropic quantum theory.  To this end we wish to list the following new aspects/results that we have obtained after doing the above analysis.

\begin{itemize}
\item
In this paper we have explicitly computed the occurrence of the pole skipping points in the complex plane for an anisotropic plasma using the corresponding dual holographic set up. We have obtained the pole skipping points from the near horizon analysis of the equation of motion for three different bulk field fluctuations: scalar, axion and the metric field perturbations. For the metric field we considered both the shear modes and the sound modes of the field fluctuations. We find that only the momentum value receives a correction due to the spatial anisotropy as parameterized by $a$ or the dimensionless ratio $b=a/T$. For scalar, axion and vector modes of the metric perturbation the pole skipping happen to appear in the lower half of the complex plane. However in the sound channel it occurs in the upper half plane and is related to the parameters of chaos. So in this regard the current paper provides a complete description of the above phenomenon for the anisotropic plasma which is one of our primary motivation.

\item As discussed earlier, the pole skipping phenomenon constraints the dispersion relation to pass through the special pole skipping points in the complex plane. In this work we have explicitly shown that the numerically obtained poles of the retarded green's function for the diffusion of transverse momentum exactly passes through the first three successive pole skipping points. Also the same kind of exact overlapping is shown to happen for the scalar field green's function.

\item  The connection between the quantum chaos and the modes of collective excitations is remarkably established by the phenomenon of pole skipping such that instead of four point functions of generic single trace operators in QFT one needs to find the points (at different orders) at which the associated energy density two point correlation function has zeroes in both the numerator and the denominator. The point at lowest order gives the butterfly velocity of quantum chaotic spread and also the Lyapunov exponent. One of the most important result of this paper is that even in the presence of a spatial anisotropy, the pole skipping phenomenon correctly produce the Lyapunov exponent and the butterfly velocity where the Lyapunov exponent saturates the chaos bound as expected and the butterfly velocity receives the anisotropic correction.

    \item  In this manuscript, for the first time (to the best of our knowledge), we have constructed the gauge invariant variables regarding the metric perturbations in both shear and scalar channel for a gravitational background dual to the anisotropic plasma. In appendix-{\bf A}, we have discussed this construction in details. Using this gauge invariant variable one can write the Einstein's equation in a simple closed form which turns out to be very useful in determining the pole skipping points.

    \item Finally, as discussed in \cite{Blake:2018leo}, a satisfactory understanding of the pole skipping phenomenon has been achieved in the gravity side in terms of a particular component of Einstein's equation which becomes trivial at the lowest order pole skipping point near the horizon. However from the field theory point of view, the reason for this phenomenon is yet to be understood. In particular, there are infinite number of special points in the complex plane where the two point correlation function has zero over zero form among which only the lowest order point has a connection to the parameters of quantum chaos. However the physical meaning of the other points is still unclear from the perspective of quantum field theory. So we hope this work would be a valuable contribution to this field of research in future.
\end{itemize}
Before closing, we must mention that in \cite{Blake:2016wvh} the author, using the gravitational shock wave analysis obtains a direct connection between the coefficient of momentum diffusion and the butterfly velocity for anisotropic background. For the special anisotropy there exists two different diffusion coefficient and hence two butterfly velocity, one along the direction of anisotropy and the other which is perpendicular to direction of anisotropy. In this paper we also compute two different butterfly velocities. Physically these two velocities indicated the speed with which the momentum diffuse in the $x_{1}$ (direction of anisotropy in our case) and $x_{2}$ direction. In \cite{Blake:2016wvh} it was shown that the ratio of these two butterfly velocities are given by the following relation,
\begin{equation}
\frac{v_{x_{1}}^2}{v_{x_{2}}^2}=\frac{g_{22}}{g_{11}}\mid_{r=r_{H}}.
\end{equation}
With the metric as given in (\ref{metric1}), the RHS of the above equation can be easily calculated as,
\begin{equation}
\frac{g_{22}}{g_{11}}|_{r=r_{H}}=1-b^2\frac{\log(2)}{4\pi^2},
\end{equation}
which exactly matches with the ratio of two velocities as obtained in (\ref{butt1}) and (\ref{butt2}) from the near horizon analysis of $vv$ component of Einstein's equation in section-{\bf 3}.

The above analysis can be repeated for a gravitational background which is deformed by the presence of uniformly distributed heavy quark in the dual field theory \cite{Chakrabortty:2011sp} \footnote{Also see \cite{Chakrabortty:2020ptb} for the study of different entanglement measures on the same back reacted background}. Moreover it would be interesting to see how the Lyapunov exponent and the butterfly velocity modify due to the non zero quark density.
\section*{Acknowledgments}
I would like to thank Shankhadeep Chakrabortty for suggesting me this problem and for regular discussions. I would also like to thank Saso Grozdanov, Xing Wu, Makoto Natsuume, Aalok Misra and Rajesh Kumar Gupta for useful discussions and comments.
\appendix
\section{Construction of the gauge invariant variable}
In this section we present a detailed discussion towards the construction of the gauge invariant variable for the anisotropic background \footnote{We find the lecture as given in \cite{winterschool} very useful in computing the gauge invariant variable in our case.}. The five dimensional metric with anisotropy (\ref{metric2}) can be rewritten following \cite{Natsuume:2019sfp} as,
\begin{equation}
ds^2=g_{ab}dx^{a}dx^{b}+g_{ij}dx^{i}dx^{j},
\end{equation}
where $x^{a}=(v,r)$ and $x^{i}=(x_{1},x_{2},x_{3})$. Note that all the metric components in the above equation depends only on the redial coordinate $r$. Also unlike $g_{ij}$, the matrixform of $g_{ab}$ has nonzero off-diagonal components. Now given the linear perturbation of the above background metric, the perturbation can be decomposed in the following way,
\begin{equation}\label{hdecom}
\begin{split}
h&=A_{1}dv\otimes dv+A_{2}dr\otimes dr+A_{3}\left(dv\otimes dr+dr\otimes dv\right)\\
&+b_{ai}\left(dx^{a}\otimes dx^{i}+dx^{i}\otimes dx^{a}\right)\\
&+\left(C g_{ij}+e_{ij}\right)dx^{i}\otimes dx^{j}.
\end{split}
\end{equation}
In the above decomposition $A_{1},A_{2},A_{3},C$ are scalars, $b_{ai}$ is a vector and $e_{ij}$ is a symmetric traceless tensor. The vector $b_{ai}$ can be decomposed as,
\begin{equation}
b_{ai}=D_{i}B_{a}+\widetilde{B}_{ai},
\end{equation}
where $B_{a}$ is a scalar for $a=(v,r)$ and $\widetilde{B}_{ai}$ is a divergence free vector, that is $D^{i}\widetilde{B}_{ai}=0$ with $D_{i}$ denoting the covariant derivative with respect to the metric $g_{ij}$. Furthermore the tensor $e_{ij}$ in (\ref{hdecom}) can also be decomposed as,
\begin{equation}
e_{ij}=\left(D_{i}D_{j}-\frac{1}{3}g_{ij}\nabla\right)E+2D_{({i}} E_{j{)}}+E_{ij},
\end{equation}
where $\nabla=g_{ij}D^{i}D^{j}$. In the above decomposition $E$ is a scalar, $E_{j}$ is a vector and $E_{ij}$ is a symmetric traceless tensor quantity.

Now consider the infinitesimal transformation of the coordinates as, $x^{m}\rightarrow x^{m}+\xi^{m}$, where $m=(a,i)$. To study the gauge invariance of the metric perturbation under the above transformation of the coordinate $x^{m}$ we first note the following definition,
\begin{equation}\label{cond}
\delta_{\xi}h_{\mu\nu}=\pounds_{\xi}g_{\mu\nu},
\end{equation}
where $\delta_{\xi}$ denotes the gauge invariant transformation and $\pounds_{\xi}$ is the lie derivative.
The infinitesimal transformation $\xi^{m}$ can again be decomposed as,
\begin{equation}
\xi^{m}=T^{a}+D^{i}L+\mathfrak{L}^{i},~~~~D_{i}\mathfrak{L}^{i}=0.
\end{equation}
\subsection{Scalar modes of metric perturbation}
The gauge invariant transformation for the scalar modes of the metric perturbation is obtained using (\ref{cond}) as,
\begin{equation}\label{gaugepert}
\begin{split}
\delta_{\xi}h_{ab}&=2D_{({a}} T_{b{)}}\\
\delta_{\xi}B_{a}&=T_{a}+\partial_{a}L-Lg^{ij}\partial_{a}g_{ij}\\
\delta_{\xi}C&=\frac{1}{3}\left(T^{a}g^{ij}\partial_{a}g_{ij}+2\nabla L\right)\\
\delta_{\xi}E&=2L.
\end{split}
\end{equation}
Substituting the final relation into the second one of the above equation we get,
\begin{equation}
\begin{split}
T_{a}&=\delta_{\xi}\zeta_{a},\\
\zeta_{a}&=B_{a}-\frac{1}{2}\partial_{a}E+\frac{1}{2}E g^{ij}\partial_{a}g_{ij}.
\end{split}
\end{equation}
Also using the above expression for $T^{a}$ in the first and the third relation of (\ref{gaugepert}) gives,
\begin{equation}\label{gaugeinv}
\begin{split}
\delta_{\xi}\mathcal{H}_{ab}&=0~~~~\rightarrow ~~~~\mathcal{H}_{ab}=h_{ab}-2D_{({a}} \zeta_{b{)}}\\
\delta_{\xi}\mathcal{H}_{L}&=0~~~~\rightarrow ~~~~~\mathcal{H}_{L}=C-\frac{1}{3}\zeta^{a}g^{ij}\partial_{a}g_{ij}-\frac{1}{3}\nabla E.
\end{split}
\end{equation}
From the above equation we see that $\mathcal{H}_{ab}$ and $\mathcal{H}_{L}$ are gauge invariant.
\subsubsection{Perturbation along the direction of the anisotropy}
For the perturbation along the direction of the anisotropy, the perturbation is written as the plane wave form: $h_{\mu\nu}=e^{-i w v+iqx_{1}}h_{\mu\nu}$. So in this case different components of the perturbation can be evaluated from (\ref{hdecom}) as,
\begin{equation}\label{diffh}
\begin{split}
h_{x_{1}x_{1}}&=Cg_{11}-q^{2}E+\left(\frac{g_{11}}{3}\right)q^{2}E\\
\frac{h_{ss}}{2}&=Cg_{22}+\frac{g_{22}}{3}q^{2}E\\
h_{vx_{1}}&=iqB_{v}.
\end{split}
\end{equation}
where we define $h_{ss}=h_{x_{2}x_{2}}+h_{x_{3}x_{3}}$, that is the trace part of the perturbation in the $(x_{2}-x_{3})$ plane.
The above equation can be solved for $C$ and $E$ to get,
\begin{equation}\label{CE}
\begin{split}
C=\frac{1}{3}h_{x_{1}x_{1}}+\frac{1}{2}g_{22}h_{ss}-\frac{1}{6}\left(\frac{g_{11}}{g_{22}}\right)h_{ss},~~~~~~~
q^{2}E =\frac{1}{2}\left(\frac{g_{11}}{g_{22}}\right)h_{ss}-h_{x_{1}x_{1}}.
\end{split}
\end{equation}
Substituting the above expression for $C$ and $q$ in (\ref{gaugeinv}) we obtain the following two gauge invariant metric perturbation for the scalar mode as,
\begin{equation}\label{gaugeinvs}
\begin{split}
\mathcal{H}_{vv}&=h_{vv}+2iw\zeta_{v}-\frac{g_{vv}^{\prime}}{g_{vr}}\left(\zeta_{v}-\frac{g_{vv}}{g_{vr}}\zeta_{r}\right)\\
\mathcal{H}_{vr}&=-\zeta_{v}^{\prime}+\left(iw+\frac{g_{vv}^{\prime}}{g_{vr}}\right)\zeta_{r}\\
\mathcal{H}_{rr}&=-\zeta_{r}^{\prime}+\frac{g_{vr}^{\prime}}{g_{vr}}\zeta_{r}\\
\mathcal{H}_{L}&=\frac{h_{ss}}{2g_{22}}-\frac{g^{ij}\partial_{r}g_{ij}}{3g_{vv}}\left(\zeta_{v}-\frac{g_{vv}^2}{g_{vr}^2}\zeta_{r}\right)
\end{split}
\end{equation}
Finally from the above equation the gauge invariant variable $Z_{s\parallel}$ for the scalar modes of metric perturbation is obtained as,
\begin{equation}\label{gaugeinvzs}
Z_{s\parallel}=2q^2\left(g_{vv}H_{vv}\right)+4wq\left(g_{11}H_{vx_{1}}\right)+2w^2\left(g_{11}H_{x_{1}x_{1}}\right)
-2\left(g_{11}w^2+3q^2\frac{g_{vv}^{\prime}}{g^{ij}\partial_{r}g_{ij}}\right)H_{x_{2}x_{2}}.
\end{equation}
The corresponding gauge invariant variable for the dilaton $(Z_{d})$ and axion $(Z_{a})$ in a similar way can be obtained as,
\begin{equation}\label{gaugeinvzda}
\begin{split}
Z_{d\parallel}&=\varphi_{0}-\frac{3\phi_{0}^{\prime}}{g^{ij}\partial_{r}g_{ij}}H_{x_{2}x_{2}},\\
Z_{a\parallel}&=\psi_{0}-a\left(\frac{i}{2q}\right)\left(H_{x_{2}x_{2}}-H_{x_{1}x_{1}}\right).
\end{split}
\end{equation}
\subsubsection{Perturbation perpendicular to the direction of the anisotropy}
Let us now take the plane wave like perturbation to propagate along the other direction say $x_{2}$ such that it can be written as, $h_{\mu\nu}=e^{-i w v+iqx_{2}}h_{\mu\nu}$. The non zero components for the scalar modes of metric perturbation are computed using (\ref{hdecom}) as,
\begin{equation}\label{hcom}
\begin{split}
h_{x_{2}x_{2}}&=C g_{22}-q^{2}E+\left(\frac{g_{22}}{3}\right)q^{2}E\\
h_{ss}&=\left(g_{11}+g_{22}\right)\left(C+\frac{1}{3}q^{2}E\right)\\
h_{vx_{2}}&=i q B_{v}.
\end{split}
\end{equation}
where in this case $h_{ss}=h_{x_{1}x_{1}}+h_{x_{3}x_{3}}$. Again the above can be solved for $C$ and $E$ as,
\begin{equation}\label{CE2}
\begin{split}
C=\frac{1}{3}h_{x_{2}x_{2}}+\frac{h_{ss}}{g_{11}+g_{22}}-\frac{g_{22}}{3(g_{11}+g_{22})}h_{ss},~~~~~
q^{2}E=\frac{g_{22}}{g_{11}+g_{22}}h_{ss}-h_{x_{2}x_{2}}.
\end{split}
\end{equation}
With this the gauge invariant metric perturbations are obtained as,
\begin{equation}\label{gaugeinvs2}
\begin{split}
\mathcal{H}_{vv}&=h_{vv}+2iw\zeta_{v}-\frac{g_{vv}^{\prime}}{g_{vr}}\left(\zeta_{v}-\frac{g_{vv}}{g_{vr}}\zeta_{r}\right)\\
\mathcal{H}_{vr}&=-\zeta_{v}^{\prime}+\left(iw+\frac{g_{vv}^{\prime}}{g_{vr}}\right)\zeta_{r}\\
\mathcal{H}_{rr}&=-\zeta_{r}^{\prime}+\frac{g_{vr}^{\prime}}{g_{vr}}\zeta_{r}\\
\mathcal{H}_{L}&=\frac{h_{ss}}{g_{11}+g_{22}}-\frac{g^{ij}\partial_{r}g_{ij}}{3g_{vv}}\left(\zeta_{v}-\frac{g_{vv}^2}{g_{vr}^2}\zeta_{r}\right)
\end{split}
\end{equation}
The gauge invariant variable $Z_{s\perp}$ is given as,
\begin{equation}\label{gaugeinvzs2}
Z_{s\perp}=2q^2\left(g_{vv}H_{vv}\right)+4wq\left(g_{22}H_{vx_{2}}\right)+2w^2\left(g_{22}H_{x_{2}x_{2}}\right)
-2\left(g_{22}w^2+3q^2\frac{g_{vv}^{\prime}}{g^{ij}\partial_{r}g_{ij}}\right)H_{x_{1}x_{1}}.
\end{equation}
Again, in this case the gauge invariant variable for the dilaton and the axion field is obtained as,
\begin{equation}\label{gaugeinvzda2}
\begin{split}
Z_{d}&=\varphi_{0}-\frac{3\phi_{0}^{\prime}}{g^{ij}\partial_{r}g_{ij}}H_{x_{1}x_{1}},\\
Z_{a}&=\psi_{0}.
\end{split}
\end{equation}
\subsection{Vector modes of metric perturbation}
For the vector modes the gauge invariant transformations are given as,
\begin{equation}\label{vec}
\begin{split}
\delta_{\xi}\tilde{B}{ai}&=\partial_{a}\mathfrak{L}_{i}-\mathfrak{L}_{i}g^{kl}\partial_{a}g_{kl},\\
\delta_{\xi}E_{i}&=\mathfrak{L}_{i}.
\end{split}
\end{equation}
Substituting the second relation into the first one in the above equation one gets,
\begin{equation}\label{gaugeinvvec}
\begin{split}
\delta_{\xi}\mathcal{H}_{ai}&=0~~~~\rightarrow ~~~~\mathcal{H}_{ai}=\tilde{B}_{ai}-\partial_{a}E_{i}+E_{i}g^{kl}\partial_{a}g_{kl}.
\end{split}
\end{equation}
Hence in this case $\mathcal{H}_{ai}$ is the gauge invariant variable.
\subsubsection{Perturbation along the direction of the anisotropy}
The nonzero components of the vector modes of metric perturbation can be expressed as,
\begin{equation}\label{vecme}
h_{vx_{2}}=\tilde{B}_{vx_{2}},~~~~~h_{x_{1}x_{2}}=iqE_{x_{2}}.
\end{equation}
So the gauge invariant variable for the vector modes with perturbation along the anisotropic direction is given as,
\begin{equation}\label{GIvec}
Z_{\mathrm{v}\parallel}=qh_{vx_{2}}+wh_{x_{1}x_{2}}.
\end{equation}
\subsubsection{Perturbation perpendicular to the direction of the anisotropy}
Again considering the perturbation along the $x_{2}$ direction with the same plane wave form the nonzero components of perturbation can be written as,
\begin{equation}\label{vecme}
h_{vx_{3}}=\tilde{B}_{vx_{3}},~~~~~h_{x_{2}x_{3}}=iqE_{x_{3}},
\end{equation}
yielding the gauge invariant variable as,
\begin{equation}\label{GIvec1}
Z_{\mathrm{v}\perp}=qh_{vx_{3}}+wh_{x_{2}x_{3}}.
\end{equation}
\section{Coefficients of the scalar field EOM}
The coefficients as appearing in equation (\ref{scleq}) and (\ref{scleq2}) for the scalar field perturbation propagating parallel and perpendicular to the direction of anisotropy are explicitly given as,
\begin{equation}\label{eq3}
\begin{split}
&S_{1}=\frac{r^3}{\left(r^4-r_{H}^4\right)}\left(-5+2i\frac{w}{r}+\frac{r_{H}^4}{r^4}\right),~~~~
S_{2}=\frac{r^2}{\left(r^4-r_{H}^4\right)}\left(\frac{q^2}{r^2}+3i\frac{w}{r}\right),\\
&\tilde{S}_{1}=\frac{r^7}{24 r_{H}^2\left(r^4-r_{H}^4\right)^2}\Biggl[6\frac{r_{H}^{6}}{r^6}-
5 i\frac{w}{r}\log \left(1+\frac {r_{H}^2} {r^2}\right)+ \frac{r_{H}^{2}} {r^2}\left(26-6 i\frac{w}{r}\right)\\&+ \frac{r_{H}^{4}} {r^4}\left\{-32+6i\frac{w}{r}-20\left(2-i\frac{w}{r}\right)\log (2) +5\left(8-3i\frac{w}{r}\right)\log\left(1+\frac{r_{H}^2}{r^2}\right)\right\}\Biggr],
\\& \tilde{S}_{2}=\frac{r^6}{48 r_{H}^2\left(r^4-r_{H}^4\right)^2}\Biggl[-\left(18\frac{q^2}{r^2}+15i\frac{w}{r}\right)
\log\left(1+\frac{r_{H}^2}{r^2}\right)+16\frac{r_{H}^2}{r^2}\left(6-\frac{q^2}{r^2}-3i\frac{w}{r}\right)\\&-96\frac{r_{H}^{6}}{r^{6}}
+\frac{r_{H}^4}{r^4}\left\{16\frac{q^2}{r^2}+48i\frac{w}{r}+20\left(\frac{q^2}{r^2}+3i\frac{w}{r}\right)\log(2)-\left(2\frac{q^2}
{r^2}+45i\frac{w}{r}\right)\log{\left(1+\frac{r_{H}^2}{r^2}\right)}\right\}\Biggr]
\\ & \tilde{\tilde{S_{1}}}=\frac{r}{24r_{H}^2\left(r^4-r_{H}^4\right)^2}
\biggl[r_{H}^2\biggl(26r^4+6r_{H}^4-6iwr^3
-8r^2r_{H}^2\left(4+5\log 2\right)+2iwr\left(3+10\log2\right)r_{H}^2\biggr)
\\&-5ir\biggl(8irr_{H}^4+r^4w+3wr_{H}^4\biggr)
\log\left(1+\frac{r_{H}^2}{r^2}\right)\biggr]\\
&\tilde{\tilde{S_{2}}}=\frac{1}{48r_{H}^2\left(r^4-r_{H}^4\right)^2}
\biggl[16r_{H}^2\left(r^4-r_{H}^4\right)
\biggl(6\left(r^2+r_{H}^2\right)-q^2-3iwr\bigg)
+20r_{H}^4\left(q^2+3iwr\right)\log2
\\&-\biggl(2q^2\left(3r^4+7r_{H}^4\right)
+15iwr\left(r^4+3r_{H}^4\right)\biggr)\log\left(1+\frac{r_{H}^2}{r^2}\right)\biggr]
\end{split}
\end{equation}
\section{Equation of motion in shear channel}
The coefficients appearing in equation (\ref{veceqn}) and (\ref{veceqn2}) are given as,
\begin{equation}\label{vecv}
\begin{split}
\mathcal{N}&=\frac{r^3 r_H^4 \left(3 r
   w^2+2 q^2 (r-i
   w)\right)-q^2 r_H^8+r^7
   \left(w^2-q^2\right) (r-2 i
   w)}{r
   \left(r^4-r_H^4\right)
   \left(r^4
   \left(q^2-w^2\right)-q^2
   r_H^4\right)}
\\\tilde{\mathcal{N}}&=\frac{1}{24rr_{H}^2\left(r^4-r_{H}^4\right)^2
\left(q^2(r^4-r_{H}^4)-w^2r^4\right)^2}
\Biggl\{2r_{H}^2q^4(r^4-r_{H}^4)^2\biggl(7r^6-12r^4r_{H}^2\\&
+3r^2r_{H}^4+2r_{H}^6-3iwr^5+iwr^3r_{H}^2(3+\log{1024})\biggr)
+2w^4r^8r_{H}^2\biggl(r^6+15r^2r_{H}^4-12r_{H}^6\\&-3iwr^5
-4r^4r_{H}^2(1+\log{32})+iwr^3r_{H}^2(3+\log{1024})\biggr)
-4q^2w^2r^4r_{H}^2\left(r^4-r_{H}^4\right)\\&\biggl(4r^6+7r^2r_{H}^4
-3iwr^5-2r^4r_{H}^2(4+\log{32})-r_{H}^6(3+\log{1024})
+iwr^3r_{H}^2(3+\log{1024})\biggr)\\&
-iwr^3\biggl(5q^4\left(r^4-r_{H}^4\right)
\left(r^4+3r_{H}^4\right)+5w^3r^8\left(8irr_{H}^4+wr^4
+3wr_{H}^4\right)-2q^2wr\left(r^4-r_{H}^4\right)\\&
\left(4ir_{H}^4\left(3r^4+7r_{H}^4\right)+5wr^3
\left(r^4+3r_{H}^4\right)\right)\biggr)
\log{\left(1+\frac{r_{H}^2}{r^2}\right)}
\Biggr\}
\end{split}
\end{equation}
\begin{equation}\label{vecv1}
\begin{split}
\mathcal{P}&=\frac{1}{r^2
\left(r^4-r_{H}^4\right)\left(q^2(r^4-r_{H}^4)-w^2r^4\right)}
\Biggl\{q^4r^2\left(r^4-r_{H}^4\right)\\&+w^2r^4
\left(-4r^4-4r_{H}^4+iwr^3\right)+q^2\left(4r^8-8r^4r_{H}^4
+4r_{H}^8-iwr^7-3iwr^3r_{H}^4-w^2r^6\right)\Biggr\}
\\\tilde{\mathcal{P}}&=\frac{1}{48r^2r_{H}^2
\left(r^4-r_{H}^4\right)^2\left(q^2(r^4-r_{H}^4)-w^2r^4\right)^2}
\Biggl\{4w^4r^8r_{H}^2\biggl(-2\left(r^2-r_{H}^2\right)\\&
\left(4r^4+12r_{H}^4-3iwr^3\right)+r^3r_{H}^2(8r-iw)\log{32}\biggr)
+4w^2q^2r^4r_{H}^2\biggl(2r^2\left(r^2-r_{H}^2\right)\\&
\biggl(2\left(7r^6-r^4r_{H}^2-7r^2r_{H}^4+r_{H}^6\right)-
iwr\left(3r^4+r^2r_{H}^2+6r_{H}^4\right)-2w^2r^4\biggr)\\&
+10r_{H}^2\left(4r_{H}^8+3iwr^7\right)\log{2}+r^3r_{H}^2
\left(-8r^5-2iwr_{H}^4+w^2r^3\right)\log{32}\biggr)\\&
-4q^6r^2r_{H}^2\left(r^4-r_{H}^4\right)^2\left(4r^2-r_{H}^2
(4+\log{32})\right)-4q^4r_{H}^2\left(r^4-r_{H}^4\right)
\biggl(20r^{10}+40r^4r_{H}^6\\&+12r^2r_{H}^8-16r_{H}^{10}
-12iwr^5r_{H}^4-8r^8(3r_{H}^2+w^2)+iwr^3r_{H}^6(14+15\log{2})\\&
+iwr^7r_{H}^2(-2+25\log{2})+2r^6r_{H}^2\left(-16r_{H}^2
+w^2(4+\log{32})\right)\biggr)
\\&-r^2\Biggl(2q^6\left(r^4-r_{H}^4\right)^2\left(9r^4+r_{H}^4\right)-iwq^4r\left(r^4-r_{H}^4\right)\biggl(5\left(r^8+22r^4r_{h}^4+9r_{H}^8\right)\\&
-4iwr^3\left(9r^4+r_{H}^4\right)\biggl)+5w^4r^9\left(32rr_{H}^4-iw\left(r^4+3r_{H}^4\right)\right)\\&+2w^2q^2r^2\biggl(16r_{H}^4\left(-3r^8-4r^4r_{H}^4+7r_{H}^8\right)
+iwr^3\left(5r^8+76r^4r_{H}^4-41r_{H}^8\right)\\&+w^2r^6\left(9r^4+r_{H}^4\right)\biggr)\Biggr)\log{\left(1+\frac{r_{H}^2}{r^2}\right)}\Biggr\}
\end{split}
\end{equation}
\begin{equation}\label{vecv3}
\begin{split}
\tilde{\tilde{\mathcal{N}}}&=\frac{1}{24rr_{H}^2\left(r^4-r_{H}^4\right)^2
\left(q^2(r^4-r_{H}^4)-w^2r^4\right)^2}
\Biggl\{2r_{H}^2q^4(r^4-r_{H}^4)^2\biggl(7r^6-12r^4r_{H}^2\\&
+3r^2r_{H}^4+2r_{H}^6-3iwr^5+iwr^3r_{H}^2(3+\log{1024})\biggr)
+2w^4r^8r_{H}^2\biggl(r^6+15r^2r_{H}^4-12r_{H}^6\\&-3iwr^5
-4r^4r_{H}^2(1+\log{32})+iwr^3r_{H}^2(3+\log{1024})\biggr)
-4q^2w^2r^4r_{H}^2\left(r^4-r_{H}^4\right)\\&\biggl(4r^6+7r^2r_{H}^4
-3iwr^5-2r^4r_{H}^2(4+\log{32})-r_{H}^6(3+\log{1024})
+iwr^3r_{H}^2(3+\log{1024})\biggr)\\&
-iwr^3\biggl(5q^4\left(r^4-r_{H}^4\right)
\left(r^4+3r_{H}^4\right)+5w^3r^8\left(8irr_{H}^4+wr^4
+3wr_{H}^4\right)-2q^2wr\left(r^4-r_{H}^4\right)\\&
\left(4ir_{H}^4\left(3r^4+7r_{H}^4\right)+5wr^3
\left(r^4+3r_{H}^4\right)\right)\biggr)
\log{\left(1+\frac{r_{H}^2}{r^2}\right)}
\Biggr\}
\end{split}
\end{equation}
\begin{equation}\label{vecv2}
\begin{split}
\tilde{\tilde{\mathcal{P}}}&=\frac{1}{48r^2r_{H}^2
\left(r^4-r_{H}^4\right)^2\left(q^2(r^4-r_{H}^4)-w^2r^4\right)^2}
\Biggl\{-4w^4r^8r_{H}^2\biggl(20r^6+12r^2r_{H}^4\\&-12r_{H}^6
+5iwr^3r_{h}^2\log{2}-20r^4r_{H}^2(1+\log{4})\biggr)-4q^6r^2r_{H}^2
\left(r^4-r_{H}^4\right)^2\left(4r^2-r_{H}^2(4+\log{32})\right)
\\&-4q^4r_{H}^2\left(r^4-r_{H}^4\right)\biggl(20r^{10}+40r^4r_{H}^6
+12r^2r_{H}^8-16r_{H}^{10}-12iwr^5r_{H}^4-8r^8\left(3r_{H}^2+w^2\right)
\\&+iwr^3r_{H}^6(14+15\log{2})+iwr^7r_{H}^2(-2+25\log{2})+2r^6r_{H}^2
\left(-16r_{H}^2+w^2(4+\log{32})\right)\biggr)\\&-4w^2q^2r^4r_{H}^2
\biggl(-40r^{10}-56r^4r_{H}^6-8r^2r_{H}^8+16iwr^5r_{H}^4
+2iwr^7r_{H}^2(1-15\log{2})+2iwr^3r_{H}^6(-9+\log{32})\\&+r^6r_{H}^2
\left(48r_{H}^2-w^2(4+\log{32})\right)-4r_{H}^{10}(-3+\log{1024})
+4r^8\left(w^2+r_{H}^2(11+\log{1024})\right)\biggr)\\&
+r^2\Biggl(-2q^6\left(r^4-r_{H}^4\right)^2
\left(3r^4+7r_{H}^4\right)+5iw^4r^9\left(32irr_{H}^4+wr^4+3wr_{H}^4\right)
\\&+wq^4r\left(r^4-r_{H}^4\right)\biggl(5i\left(r^8+22r^4r_{H}^4+9r_{H}^8\right)
+4wr^3\left(3r^4+7r_{H}^4\right)\biggr)+2w^2q^2r^2\\&
\biggl(-32r_{H}^4\left(-3r^8+r^4r_{H}^4+2r_{H}^8\right)-iwr^3
\left(5r^8+52r^4r_{H}^4-17r_{H}^8\right)\\&-r^6w^2
\left(3r^4+7r_{H}^4\right)\biggr)\Biggl)\log{\left(1+\frac{r_{H}^2}{r^2}\right)}
\Biggr\}
\end{split}
\end{equation}
The coefficients appearing in equation (\ref{difusion}) are given as,
\begin{equation}\label{difucoff}
\begin{split}
&\mathcal{R}=\frac{Q^2(1-u^2)^2-(1+u^2)W^2}{u(1-u^2)(W^2-Q^2(1-u^2))}
,~~~~~~~~~~~\mathcal{S}=\frac{Q(1-u^2)-W^2}{u(1-u)^2},\\
&\tilde{\mathcal{R}}=\frac{1}{24\pi^2(1-u^2)(1+u)^2(W^2-Q^2(1-u^2))^2}
\biggl\{-5Q^4(1-u)^4(1+u)^2(5+4u)\\&+W^4\biggl(-19+22u+3u^2-6u^3+20u\log{(2)}\biggr)
-4uW^2\biggl(Q^2(-3-4u^2+7u^4)+5W^2\biggr)\\&\times\log{(1+u)}-2Q^2W^2(1-u^2)\biggl(-22+26u+11u^2+5u^2(-3+\log{(4)})+u\log{(1024)}\biggr)
\biggr\}\\&
\tilde{\mathcal{S}}=\frac{1}{24\pi^2u(1-u^2)^2(u^2-1)(W^2-Q^2(1-u^2))^2}
\biggl\{W^2\biggl(-6-6u^4-6uW^2+12u^2\\&+u^2W^2(8+15\log{(2)})
+W^2(-2+\log(32))\biggr)+Q^4(1-u^2)^2\biggl(-2-8u\\&+5u^2(2+\log(2))
+\log(32)\biggr)-2Q^2(1-u^2)\biggl(-3-6u^3-9u^4+6u^5-7uW^2\\&+
W^2(-2+\log(32))+12u^2+u^2W^2(9+\log(1024))\biggr)\\&-\log(1+u)\biggl(Q^4
(1-u^2)^2(9+u^2)+2Q^2W^2(-7-u^2+8u^4)+5W^4(1+3u^2)\biggr)\biggr\},
\end{split}
\end{equation}
\section{Equation of motion in sound channel}
The coefficients of the equations in (\ref{GIqnn}) are given as,
\begin{equation}\label{coeff}
\begin{split}
&\mathcal{M}=\frac{3r^4w^2\left(r^4+3r_{H}^4-2iwr^{3}\right)-q^2\left(3r^8+5r_{H}^8-6iwr^{7}+2ir^{3}r_{H}^4w\right)}{r\left(r^4-r_{H}^4\right)
\left(q^2\left(3r^4-r_{H}^4\right)-3w^2r^4\right)}
\end{split}
\end{equation}
\begin{equation}\label{coeff2}
\begin{split}
\mathcal{L}&=\frac{1}{r^2\left(r^4-r_{H}^4\right)
\left(q^2\left(3r^4-r_{H}^4\right)-3r^4w^2\right)}\Biggl\{q^4r^2\left(3r^4-r_{H}^{4}\right)\\&-3w^2r^4\left(4r^4+4r_{H}^{4}-iwr^3\right)
+q^2\left(12r^8-8r^4r_{H}^{4}-4r_{H}^{8}-3iwr^7-7iwr^3r_{H}^{4}-3w^2r^6\right)\Biggr\}
\end{split}
\end{equation}
\begin{equation}\label{coeff1}
\begin{split}
\tilde{\mathcal{M}}&=\frac{1}{24 r r_H^2 \left(r^2-r_H^2\right){}^2
   \left(r_H^2+r^2\right){}^2 \left(q^2 r_H^4+3 r^4
   \left(w^2-q^2\right)\right){}^2}\Biggl\{2r_{H}^2\biggl(45 q^4 r^{14}\\&
  -27 i q^4 r^{13} w+36 i q^2 r^{15} w-54 q^2 r^{14}
   w^2+54 i q^2 r^{13} w^3-108 i r^{15} w^3-9
   r^{14} w^4\\&-84 q^4 r^{12} r_H^2+27 i q^4 r^{11} w r_H^2+120
   q^2 r^{12} w^2 r_H^2-54 i q^2 r^{11} w^3
   r_H^2-18 r^{12} w^4 r_H^2\\&-27 i r^{13} w^5120 q^2 r^{12} w^2 \log (4) r_H^2-180 r^{12} w^4
   \log (2) r_H^2+27 i r^{11} w^5 r_H^2+90 i r^{11}
   w^5 \log (2) r_H^2\\&-12 q^4 r^{12} \log (32) r_H^2+9 i q^4 r^{11} w
   \log (1024) r_H^2+57 q^4 r^{10} r_H^4-18 i q^2
   r^{11} w^3 \log (1024) r_H^2\\&+18 i q^4 r^9 w r_H^4-36 i q^2 r^{11} w r_H^4-192
   q^2 r^{10} w^2 r_H^4-18 i q^2 r^9 w^3 r_H^4+216
   i r^{11} w^3 r_H^4+153 r^{10} w^4 r_H^4\\&-120 q^4 r^8 \log (2) r_H^6-18 i q^4 r^7 w
   r_H^6+120 q^2 r^8 w^2 r_H^6+18 i q^2 r^7 w^3
   r_H^6-126 r^8 w^4 r_H^6\\&-6 i q^4 r^7 w \log (1024) r_H^6+3 q^4 r^6
   r_H^8+120 q^2 r^8 w^2 \log (2) r_H^6+6 i q^2 r^7
   w^3 \log (1024) r_H^6\\&-3 i q^4 r^5 w r_H^8-28 q^4 r^4 r_H^{10}+3 i q^4
   r^3 w r_H^{10}-36 i q^2 r^7 w r_H^8+6 q^2 r^6
   w^2 r_H^8-108 i r^7 w^3 r_H^8\\&100 q^4 r^4 \log (2) r_H^{10}+i q^4 r^3 w \log
   (1024) r_H^{10}-9 q^4 r^2 r_H^{12}-120 q^2 r^4
   w^2 \log (2) r_H^{10}+16 q^4 r_H^{14}\\&+36 i q^2 r^3 w r_H^{12}\biggr)+\log{\left(1+\frac{r_{H}^2}{r^2}\right)}
   \biggl(18 q^4 r^{16}-45 i q^4 r^{15} w+36 q^2 r^{16} w^2\\&+12 q^4 r^{12} r_H^4-105 i q^4 r^{11} w r_H^4-420
   q^2 r^{12} w^2 r_H^4+90 i q^2 r^{15} w^3-45 i
   r^{15} w^5\\&+408 q^4 r^8 r_H^8+85 i q^4 r^7 w r_H^8+240 i q^2
   r^{11} w^3 r_H^4+360 r^{12} w^4 r_H^4-135 i
   r^{11} w^5 r_H^4 \\&-284 q^4 r^4 r_H^{12}-15 i q^4 r^3 w r_H^{12}+6 q^4
   r_H^{16}-468 q^2 r^8 w^2 r_H^8-90 i q^2 r^7 w^3
   r_H^8+372 q^2 r^4 w^2 r_H^{12} \biggr)\Biggr\}
\end{split}
\end{equation}
\begin{equation}\label{coeff3}
\begin{split}
\nonumber \tilde{\mathcal{L}}&=\frac{1}{48 q^4 r^2 r_H^2 \left(r^4-r_H^4\right){}^4 \left(3
r^4-r_H^4\right) \left(q^2 r_H^4+3 r^4
\left(w^2-q^2\right)\right){}^2}\Biggl\{r_{H}^2\biggl(-432 q^{10} r^{24}\\&-1728 q^8 r^{26}-864 i q^6 r^{27}
   w +756 q^8 r^{24} w^2+108 q^6 r^{26} w^2+648 i q^6
   r^{25} w^3+2592 i q^4 r^{27} w^3\\&-108 q^6 r^{24} w^4+4104 q^4 r^{26} w^4-1080 i q^4
   r^{25} w^5-216 q^4 r^{24} w^6-2916 q^2 r^{26}
   w^6+648 i q^2 r^{25} w^7\\&+432 q^{10} r^{22} r_H^2+2664 q^8 r^{24} r_H^2-72 i
   q^8 r^{23} w r_H^2-756 q^8 r^{22} w^2 r_H^2-4500
   q^6 r^{24} w^2 r_H^2\\&+72 i q^6 r^{23} w^3 r_H^2+108 q^6 r^{22} w^4 r_H^2+1728 q^4 r^{24} w^4
   r_H^2+432 i q^4 r^{23} w^5 r_H^2+216 q^4 r^{22}
   w^6 r_H^2\\&-324 q^2 r^{24} w^6 r_H^2-648 i q^2
   r^{23} w^7 r_H^2 +1440 q^8 r^{24} \log (2) r_H^2-1980 i q^8 r^{23} w
   \log (2) r_H^2\\&-5760 q^6 r^{24} w^2 \log (2) r_H^2+4320 q^4 r^{24}
   w^4 \log (2) r_H^2-540 i q^4 r^{23} w^5 \log (2)
   r_H^2+108 q^{10} r^{22} \log (32) r_H^2\\&+360 i q^6 r^{23}
   w^3 \log (128) r_H^2+180 q^6 r^{22} w^4 \log (8)
   r_H^2+1296 q^{10} r^{20} r_H^4-108 q^8 r^{22} w^2 \log
   (1024) r_H^2\\&+3312 q^8 r^{22} r_H^4+288 i q^8
   r^{21} w r_H^4+720 i q^6 r^{23} w r_H^4-2016 q^8 r^{20} w^2 r_H^4+9108 q^6 r^{22} w^2
   r_H^4\\&-1728 i q^6 r^{21} w^3 r_H^4+396 q^6 r^{20}
   w^4 r_H^4-3456 i q^4 r^{23} w^3 r_H^4-18216 q^4 r^{22} w^4 r_H^4+1656 i q^4 r^{21} w^5
   r_H^4\\&+288 q^4 r^{20} w^6 r_H^4-3888 i q^2 r^{23}
   w^5 r_H^4+8100 q^2 r^{22} w^6 r_H^4+2520 i q^6 r^{21} w^3 \log (2) r_H^4-180 q^6 r^{20}
   w^4 \log (8) r_H^4\\&+540 q^6 r^{20} w^4 \log (2)
   r_H^4-864 i q^2 r^{21} w^7 r_H^4-1296 q^{10} r^{18} r_H^6-4992 q^8 r^{20} r_H^6-360
   i q^6 r^{21} w^3 \log (128) r_H^4\\&-264 i q^8 r^{19} w r_H^6+2016 q^8 r^{18} w^2
   r_H^6+1860 q^6 r^{20} w^2 r_H^6-96 i q^6 r^{19}
   w^3 r_H^6-396 q^6 r^{18} w^4 r_H^6\\&+4392 q^4 r^{20} w^4 r_H^6-144 i q^4 r^{19} w^5
   r_H^6-288 q^4 r^{18} w^6 r_H^6-540 q^2 r^{20}
   w^6 r_H^6+864 i q^2 r^{19} w^7 r_H^6\\&-6240 q^8 r^{20} \log (2) r_H^6+4980 i q^8 r^{19} w
   \log (2) r_H^6+13440 q^6 r^{20} w^2 \log (2)
   r_H^6\\&-2520 i q^6 r^{19} w^3 \log (2) r_H^6-540 q^6
   r^{18} w^4 \log (2) r_H^6-10080 q^4 r^{20} w^4
   \log (2) r_H^6\\&+720 q^6 r^{20} w^2 \log (16) r_H^6-240 q^6 r^{18}
   w^4 \log (8) r_H^6+1260 i q^4 r^{19} w^5 \log
   (2) r_H^6\\&-324 q^{10} r^{18} \log (32) r_H^6-480 i q^6 r^{19}
   w^3 \log (128) r_H^6-72 i q^6 r^{19} w^3 \log
   (32) r_H^6\\&-1440 q^{10} r^{16} r_H^8+288 q^8 r^{18} w^2 \log
   (1024) r_H^6-1488 q^8 r^{18} r_H^8-1344 i q^8
   r^{17} w r_H^8\\&+1848 q^8 r^{16} w^2 r_H^8+1584 i q^6 r^{19} w
   r_H^8-22200 q^6 r^{18} w^2 r_H^8+2448 i q^6
   r^{17} w^3 r_H^8\\&-372 q^6 r^{16} w^4 r_H^8-1728 i q^4 r^{19} w^3
   r_H^8+24408 q^4 r^{18} w^4 r_H^8-648 i q^4
   r^{17} w^5 r_H^8+9072 i q^2 r^{19} w^5 r_H^8\\&+2880 q^6 r^{18} w^2 \log (2) r_H^8-72 q^4 r^{16}
   w^6 r_H^8-6588 q^2 r^{18} w^6 r_H^8+216 i q^2
   r^{17} w^7 r_H^8\\&-3360 i q^6 r^{17} w^3 \log (2) r_H^8+240 q^6
   r^{16} w^4 \log (8) r_H^8-720 q^6 r^{16} w^4
   \log (2) r_H^8\\&+1440 q^{10} r^{14} r_H^{10}+1224 q^8 r^{16}
   r_H^{10}-720 q^6 r^{18} w^2 \log (16) r_H^8+480
   i q^6 r^{17} w^3 \log (128) r_H^8\\&+1680 i q^8 r^{15} w r_H^{10}-1848 q^8 r^{14} w^2
   r_H^{10}+13128 q^6 r^{16} w^2 r_H^{10}-912 i q^6
   r^{15} w^3 r_H^{10}\\&+372 q^6 r^{14} w^4 r_H^{10}-14184 q^4 r^{16} w^4
   r_H^{10}-432 i q^4 r^{15} w^5 r_H^{10}+72 q^4
   r^{14} w^6 r_H^{10}+1188 q^2 r^{16} w^6 r_H^{10}\\&+10560 q^8 r^{16} \log (2) r_H^{10}-3720 i q^8
   r^{15} w \log (2) r_H^{10}-216 i q^2 r^{15} w^7
   r_H^{10}\\&-12480 q^6 r^{16} w^2 \log (2) r_H^{10}+3360 i q^6
   r^{15} w^3 \log (2) r_H^{10}+720 q^6 r^{14} w^4
   \log (2) r_H^{10}\\&+60 q^6 r^{14} w^4 \log (8) r_H^{10}+7200 q^4 r^{16}
   w^4 \log (2) r_H^{10}-900 i q^4 r^{15} w^5 \log
   (2) r_H^{10}\\&+360 q^{10} r^{14} \log (32) r_H^{10}-960 q^6 r^{16}
   w^2 \log (16) r_H^{10}+168 i q^6 r^{15} w^3 \log
   (32) r_H^{10}\\&+736 q^{10} r^{12} r_H^{12}-264 q^8 r^{14} w^2 \log
   (1024) r_H^{10}-864 q^8 r^{14} r_H^{12}+120 i
   q^6 r^{15} w^3 \log (128) r_H^{10}\\&+1664 i q^8 r^{13} w r_H^{12}-672 q^8 r^{12} w^2
   r_H^{12}-1440 i q^6 r^{15} w r_H^{12}+19128 q^6
   r^{14} w^2 r_H^{12}\\&-1728 i q^6 r^{13} w^3 r_H^{12}+84 q^6 r^{12} w^4
   r_H^{12}+3456 i q^4 r^{15} w^3 r_H^{12}-12312
   q^4 r^{14} w^4 r_H^{12}\\&-3840 q^6 r^{14} w^2 \log (2) r_H^{12}+72 i q^4
   r^{13} w^5 r_H^{12}-6480 i q^2 r^{15} w^5
   r_H^{12}+1404 q^2 r^{14} w^6 r_H^{12}\\&+840 i q^6 r^{13} w^3 \log (2) r_H^{12}-60 q^6
   r^{12} w^4 \log (8) r_H^{12}+180 q^6 r^{12} w^4
 \log (2) r_H^{12}
\end{split}
\end{equation}
\begin{equation}
\begin{split}
&-736 q^{10} r^{10} r_H^{14}+3072 q^8 r^{12}
   r_H^{14}+960 q^6 r^{14} w^2 \log (16)
   r_H^{12}-120 i q^6 r^{13} w^3 \log (128)
   r_H^{12}\\&-2160 i q^8 r^{11} w r_H^{14}+672 q^8 r^{10} w^2
   r_H^{14}-16296 q^6 r^{12} w^2 r_H^{14}+1248 i
   q^6 r^{11} w^3 r_H^{14}-84 q^6 r^{10} w^4
   r_H^{14}\\&-8640 q^8 r^{12} \log (2) r_H^{14}+10008 q^4 r^{12}
   w^4 r_H^{14}+144 i q^4 r^{11} w^5 r_H^{14}-324
   q^2 r^{12} w^6 r_H^{14}\\&+760 i q^8 r^{11} w \log (2) r_H^{14}+5760 q^6
   r^{12} w^2 \log (2) r_H^{14}-840 i q^6 r^{11}
   w^3 \log (2) r_H^{14}\\&-180 q^6 r^{10} w^4 \log (2) r_H^{14}-1440 q^4
   r^{12} w^4 \log (2) r_H^{14}+180 i q^4 r^{11}
   w^5 \log (2) r_H^{14}\\&-184 q^{10} r^{10} \log (32) r_H^{14}-96 i q^8
   r^{11} w \log (32) r_H^{14}+240 q^6 r^{12} w^2
   \log (16) r_H^{14}\\&-176 q^{10} r^8 r_H^{16}+96 q^8 r^{10} w^2 \log
   (1024) r_H^{14}+1120 q^8 r^{10} r_H^{16}-120 i
   q^6 r^{11} w^3 \log (32) r_H^{14}\\&-704 i q^8 r^9 w r_H^{16}+84 q^8 r^8 w^2
   r_H^{16}-576 i q^6 r^{11} w r_H^{16}-7092 q^6
   r^{10} w^2 r_H^{16}+360 i q^6 r^9 w^3 r_H^{16}\\&+960 q^6 r^{10} w^2 \log (2) r_H^{16}-864 i q^4
   r^{11} w^3 r_H^{16}+2016 q^4 r^{10} w^4
   r_H^{16}+1296 i q^2 r^{11} w^5 r_H^{16}\\&+176 q^{10} r^6 r_H^{18}-2760 q^8 r^8 r_H^{18}+952 i
   q^8 r^7 w r_H^{18}-84 q^8 r^6 w^2 r_H^{18}-240
   q^6 r^{10} w^2 \log (16) r_H^{16}\\&+3360 q^8 r^8 \log (2) r_H^{18}+6732 q^6 r^8 w^2
   r_H^{18}-312 i q^6 r^7 w^3 r_H^{18}-1944 q^4 r^8
   w^4 r_H^{18}\\&+44 q^{10} r^6 \log (32) r_H^{18}+420 i q^8 r^7 w
   \log (2) r_H^{18}-960 q^6 r^8 w^2 \log (2)
   r_H^{18}\\&+32 i q^8 r^7 w \log (32) r_H^{18}-12 q^8 r^6 w^2
   \log (1024) r_H^{18}+24 i q^6 r^7 w^3 \log (32)
   r_H^{18}\\&+16 q^{10} r^4 r_H^{20}-16 q^{10} r^2 r_H^{22}-400
   q^8 r^6 r_H^{20}+96 i q^8 r^5 w r_H^{20}+720 i
   q^6 r^7 w r_H^{20}+948 q^6 r^6 w^2 r_H^{20}\\&+896 q^8 r^4 r_H^{22}-480 q^8 r^4 \log (2)
   r_H^{22}-136 i q^8 r^3 w r_H^{22}-924 q^6 r^4
   w^2 r_H^{22}\\&-4 q^{10} r^2 \log (32) r_H^{22}-140 i q^8 r^3 w
   \log (2) r_H^{22}+48 q^8 r^2 r_H^{24}-104 q^8
   r_H^{26}-144 i q^6 r^3 w r_H^{24}\biggr)\\&+\biggl(-432 q^{10} r^{26}+270 q^8 r^{28}-297 i q^8 r^{27}w+918 q^8 r^{26} w^2-432 q^6 r^{28} w^2-486 iq^6 r^{27} w^3-702 q^6 r^{26} w^4\\&+1260 q^{10} r^{22} r_H^4-1458 q^4 r^{28} w^4+1107 i q^4 r^{27} w^5+216 q^4 r^{26} w^6+972 q^2 r^{28}w^6-648 i q^2 r^{27} w^7\\&-2718 q^8 r^{24} r_H^4+3240 i q^8 r^{23} wr_H^4-2574 q^8 r^{22} w^2 r_H^4+6156 q^6 r^{24}w^2 r_H^4-1098 i q^6 r^{23} w^3 r_H^4\\&+1872 q^6 r^{22} w^4 r_H^4-54 q^4 r^{24} w^4r_H^4-2934 i q^4 r^{23} w^5 r_H^4-504 q^4 r^{22}w^6 r_H^4-1944 q^2 r^{24} w^6 r_H^4\\&-1380 q^{10} r^{18} r_H^8+9030 q^8 r^{20}r_H^8-7461 i q^8 r^{19} w r_H^8+2628 q^8 r^{18}w^2 r_H^8+1512 i q^2 r^{23} w^7 r_H^4\\&-15900 q^6 r^{20} w^2 r_H^8+5532 i q^6 r^{19} w^3r_H^8-1716 q^6 r^{18} w^4 r_H^8+6660 q^4 r^{20}w^4 r_H^8+2664 i q^4 r^{19} w^5 r_H^8\\&+744 q^{10} r^{14} r_H^{12}-14166 q^8 r^{16}r_H^{12}+360 q^4 r^{18} w^6 r_H^8+864 q^2 r^{20}w^6 r_H^8-1080 i q^2 r^{19} w^7 r_H^8\\&+6416 i q^8 r^{15} w r_H^{12}-1236 q^8 r^{14} w^2 r_H^{12}+16584 q^6 r^{16} w^2 r_H^{12}-6204 iq^6 r^{15} w^3 r_H^{12}\\&+624 q^6 r^{14} w^4 r_H^{12}-7668 q^4 r^{16} w^4r_H^{12}-954 i q^4 r^{15} w^5 r_H^{12}-72 q^4r^{14} w^6 r_H^{12}+216 q^2 r^{16} w^6 r_H^{12}\\&-232 q^{10} r^{10} r_H^{16}+11466 q^8 r^{12}r_H^{16}-1819 i q^8 r^{11} w r_H^{16}+294 q^8r^{10} w^2 r_H^{16}+216 i q^2 r^{15} w^7 r_H^{12}\\&-8088 q^6 r^{12} w^2 r_H^{16}+2634 i q^6 r^{11} w^3 r_H^{16}-78 q^6 r^{10} w^4r_H^{16}+2862 q^4r^{12} w^4 r_H^{16}+117 i q^4 r^{11} w^5r_H^{16}\\&+44 q^{10} r^6 r_H^{20}-4602 q^8 r^8 r_H^{20}-184 iq^8 r^7 w r_H^{20}-30 q^8 r^6 w^2 r_H^{20}-108q^2 r^{12} w^6 r_H^{16}\\&-4 q^{10} r^2 r_H^{24}+738 q^8 r^4 r_H^{24}+1836q^6 r^8 w^2 r_H^{20}-378 i q^6 r^7 w^3r_H^{20}-342 q^4 r^8 w^4 r_H^{20}\\&+105 i q^8 r^3 w r_H^{24}-18 q^8 r_H^{28}-156 q^6r^4 w^2 r_H^{24}
\biggr)\log{\left(1+\frac{r_{H}^2}{r^2}\right)}\Biggr\}
\end{split}
\end{equation}
\begin{equation}\label{coeff4}
\begin{split}
\tilde{\tilde{\mathcal{M}}}&=\frac{1}{24 q^2 r_H^2 \left(r^2-r_H^2\right){}^2
   \left(r_H^2+r^2\right){}^2 \left(3 r^5-r
   r_H^4\right) \left(q^2 r_H^4+3 r^4
   \left(w^2-q^2\right)\right){}^2}\Biggl\{r_{H}^2\biggl(A_{1}w^6\\&+A_{2}q^2w^4
+A_{3}q^4w^2+A_{4}q^6\biggr)
+q^2\left(3r^4-r_{H}^4\right)\biggl(A_{5}w^4+A_{6}q^4+A_{7}q^2w^2\biggr)
\log{\left(1+\frac{r_{H}^2}{r^2}\right)}\Biggr\}
\end{split}
\end{equation}
with
\begin{equation}\label{co}
\begin{split}
A_{1}&=-54r^{14}\left(r^4-r_{H}^4\right)\\
A_{2}&=18r^8\biggl(6r^9\left(3r-iw\right)+6r^7r_{H}^2\biggl\{
-5r\left(1+\log{4}\right)+iw\left(1+\log{32}\right)\biggr\}
+r^5r_{H}^4\left(39r+2iw\right)\\&+r^3r_{H}^6\biggl\{5r\left(-5+\log{16}\right)
-2iw\left(1+\log{32}\right)\biggr\}-13r^2r_{H}^8+11r_{H}^{10}biggr)\\
A_{3}&=-3r^4\left(3r^4-r_{H}^4\right)\biggl(3r^9\left(9r-10iw\right)
+r^7r_{H}^2\biggl\{-r\left(83+160\log{2}\right)+30iw\left(1+\log{16}\right)\biggr\}
\\&+2r^5r_{H}^4\left(62r+5iw\right)-2r^3r_{H}^6\biggl\{5iw\left(1+
\log{16}\right)+8r\left(4+\log{32}\right)\biggr\}\\&+33r^2r_{H}^8+
r_{H}^{10}\left(-37+80\log{2}\right)\biggr)\\
A_{4}&=\left(-3r^4+r_{H}^4\right)\biggl(9r^{13}\left(r+6iw\right)+
3r^{11}r_{H}^2\biggl\{r\left(23+40\log{2}\right)-6iw\left(3+10\log{2}\right)
\biggr\}\\&-3r^9r_{H}^4\left(43r+12iw\right)+3r^7r_{H}^6\biggl\{
r\left(5+80\log{2}\right)+4iw\left(3+10\log{2}\right)\biggr\}+
3r^5r_{H}^8\left(-35r+2iw\right)\\&-r^3r_{H}^{10}\biggl\{5r
\left(-31+40\log{2}\right)+2iw\left(3+10\log{2}\right)\biggr\}+
33r^2r_{H}^{12}-47r_{H}^{14}\biggr)\\
A_{5}&=45r^{11}\biggl(8rr_{H}^4-iw\left(r^4+3r_{H}^4\right)\biggr)\\
A_{6}&=54r^{16}-204r^{12}r_{H}^4+744r^8r_{H}^8-452r^4r_{H}^{12}+
18r_{H}^{16}-5iwr^3\left(-3r^4+r_{H}^4\right)^2\left(r^4+3r_{H}^4\right)
\end{split}
\end{equation}
\begin{equation}\label{coeff5}
\begin{split}
\nonumber \tilde{\tilde{\mathcal{L}}}&=\frac{1}{48 q^4 r^2 r_H^2 \left(r^4-r_H^4\right){}^4
   \left(r_H^4-3 r^4\right){}^2 \left(q^2 \left(3
   r^4-r_H^4\right)-3 r^4 w^2\right){}^2}\Biggl\{r_{H}^2\biggl(-1539 q^{10} r^{28}\\&-6723 q^8 r^{30}+3888 i q^8
   r^{29} w +3564 q^8 r^{28} w^2+18711 q^6 r^{30} w^2-6804 i q^6
   r^{29} w^3\\&-3159 q^6 r^{28} w^4+162 i q^6 r^{27} w^5-12879 q^4
   r^{30} w^4+4050 i q^4 r^{29} w^5+1944 q^4 r^{28}
   w^6\\&-324 i q^4 r^{27} w^7-567 q^2 r^{30} w^6-1134 i q^2
   r^{29} w^7-1296 q^2 r^{28} w^8+1458 r^{30} w^8\\&+1539 q^{10} r^{26} r_H^2+4671 q^8 r^{28} r_H^2-3456
   i q^8 r^{27} w r_H^2+162 i q^2 r^{27} w^9+486
   r^{28} w^{10}\\&-3402 q^8 r^{26} w^2 r_H^2-13581 q^6 r^{28} w^2
   r_H^2+2808 i q^6 r^{27} w^3 r_H^2+2187 q^6
   r^{26} w^4 r_H^2\\&+9153 q^4 r^{28} w^4 r_H^2-162 i q^6 r^{25} w^5 r_H^2+5022 i q^4 r^{27} w^5
   r_H^2-486 q^4 r^{26} w^6 r_H^2\\&+324 i q^4 r^{25} w^7 r_H^2-243 q^2 r^{28} w^6
   r_H^2-7290 i q^2 r^{27} w^7 r_H^2+648 q^2 r^{26}
   w^8 r_H^2\\&+1620 q^{10} r^{26} \log (2) r_H^2-162 i q^2 r^{25}
   w^9 r_H^2+2916 i r^{27} w^9 r_H^2-486 r^{26}
   w^{10} r_H^2\\&+4320 q^8 r^{28} \log (2) r_H^2-5940 i q^8 r^{27} w
   \log (2) r_H^2-3240 q^8 r^{26} w^2 \log (2)
   r_H^2\\&-17280 q^6 r^{28} w^2 \log (2) r_H^2+7560 i q^6
   r^{27} w^3 \log (2) r_H^2+1620 q^6 r^{26} w^4
   \log (2) r_H^2\\&+4806 q^{10} r^{24} r_H^4+12960 q^4 r^{28} w^4 \log
   (2) r_H^2-162 i q^4 r^{27} w^5 \log (1024) r_H^2\\&+31833 q^8 r^{26} r_H^4-5616 i q^8 r^{25} w
   r_H^4-9558 q^8 r^{24} w^2 r_H^4-62424 q^6 r^{26}
   w^2 r_H^4\\&+16092 i q^6 r^{25} w^3 r_H^4+7668 q^6 r^{24} w^4
   r_H^4-162 i q^6 r^{23} w^5 r_H^4+26001 q^4
   r^{26} w^4 r_H^4\\&-21816 i q^4 r^{25} w^5 r_H^4-4374 q^4 r^{24} w^6
   r_H^4+216 i q^4 r^{23} w^7 r_H^4+9990 q^2 r^{26}
   w^6 r_H^4\\&+14310 i q^2 r^{25} w^7 r_H^4+1728 q^2 r^{24} w^8
   r_H^4-54 i q^2 r^{23} w^9 r_H^4-6804 r^{26} w^8
   r_H^4\\&-4806 q^{10} r^{22} r_H^6-26469 q^8 r^{24}
   r_H^6-3888 i r^{25} w^9 r_H^4-162 r^{24} w^{10}
   r_H^4\\&+4464 i q^8 r^{23} w r_H^6+9072 q^8 r^{22} w^2
   r_H^6+45072 q^6 r^{24} w^2 r_H^6-6840 i q^6
   r^{23} w^3 r_H^6\\&-5508 q^6 r^{22} w^4 r_H^6+162 i q^6 r^{21} w^5
   r_H^6-6399 q^4 r^{24} w^4 r_H^6+3672 i q^4
   r^{23} w^5 r_H^6\\&+1836 q^4 r^{22} w^6 r_H^6-216 i q^4 r^{21} w^7
   r_H^6-16362 q^2 r^{24} w^6 r_H^6-486 i q^2
   r^{23} w^7 r_H^6\\&-5400 q^{10} r^{22} \log (2) r_H^6-864 q^2 r^{22}
   w^8 r_H^6+54 i q^2 r^{21} w^9 r_H^6+5346 r^{24}
   w^8 r_H^6\\&+162 r^{22} w^{10} r_H^6-20160 q^8 r^{24} \log (2) r_H^6+3960 i q^8 r^{23}
   w \log (2) r_H^6\\&+2160 q^8 r^{22} w^2 \log (2) r_H^6+54720 q^6 r^{24}
   w^2 \log (2) r_H^6-21240 i q^6 r^{23} w^3 \log
   (2) r_H^6\\&-1080 q^6 r^{22} w^4 \log (2) r_H^6-34560 q^4
   r^{24} w^4 \log (2) r_H^6+3780 i q^4 r^{23} w^5
   \log (2) r_H^6\\&+2160 i q^8 r^{23} w \log (64) r_H^6+756 q^8 r^{22}
   w^2 \log (1024) r_H^6-810 q^6 r^{22} w^4 \log
   (16) r_H^6\\&-5697 q^{10} r^{20} r_H^8-48735 q^8 r^{22}
   r_H^8-4536 i q^8 r^{21} w r_H^8+54 i q^4 r^{23}
   w^5 \log (1024) r_H^6\\&+9396 q^8 r^{20} w^2 r_H^8+77715 q^6 r^{22} w^2
   r_H^8-5760 i q^6 r^{21} w^3 r_H^8-6030 q^6
   r^{20} w^4 r_H^8\\&+54 i q^6 r^{19} w^5 r_H^8-36558 q^4 r^{22} w^4
   r_H^8+18972 i q^4 r^{21} w^5 r_H^8+2376 q^4
   r^{20} w^6 r_H^8\\&-36 i q^4 r^{19} w^7 r_H^8+6696 q^2 r^{22} w^6
   r_H^8-9018 i q^2 r^{21} w^7 r_H^8-432 q^2 r^{20}
   w^8 r_H^8\\&+5697 q^{10} r^{18} r_H^{10}+44763 q^8 r^{20}
   r_H^{10}+1134 r^{22} w^8 r_H^8+1296 i r^{21} w^9
   r_H^8\\&+5736 i q^8 r^{19} w r_H^{10}-8856 q^8 r^{18} w^2
   r_H^{10}-58569 q^6 r^{20} w^2 r_H^{10}-1560 i
   q^6 r^{19} w^3 r_H^{10}\\&+4374 q^6 r^{18} w^4 r_H^{10}-54 i q^6 r^{17} w^5
   r_H^{10}+10890 q^4 r^{20} w^4 r_H^{10}-7020 i
   q^4 r^{19} w^5 r_H^{10}\\&-1098 q^4 r^{18} w^6 r_H^{10}+36 i q^4 r^{17} w^7
   r_H^{10}+4752 q^2 r^{20} w^6 r_H^{10}+2754 i q^2
   r^{19} w^7 r_H^{10}\\&+7020 q^{10} r^{18} \log (2) r_H^{10}+216 q^2 r^{18}
   w^8 r_H^{10}-1296 r^{20} w^8 r_H^{10}-324 i
   r^{19} w^9 r_H^{10}\\&+37920 q^8 r^{20} \log (2) r_H^{10}-7500 i q^8
   r^{19} w \log (2) r_H^{10}-5760 q^8 r^{18} w^2
   \log (2) r_H^{10}
\end{split}
\end{equation}
\begin{equation}
\begin{split}
\nonumber &-65280 q^6 r^{20} w^2 \log (2) r_H^{10}+21360 i q^6
   r^{19} w^3 \log (2) r_H^{10}+1800 q^6 r^{18} w^4
   \log (2) r_H^{10}\\&+540 q^6 r^{18} w^4 \log (16) r_H^{10}+31680 q^4
   r^{20} w^4 \log (2) r_H^{10}-3960 i q^4 r^{19}
   w^5 \log (2) r_H^{10}\\&+3252 q^{10} r^{16} r_H^{12}-1440 i q^8 r^{19} w
   \log (64) r_H^{10}-504 q^8 r^{18} w^2 \log
   (1024) r_H^{10}\\&+28941 q^8 r^{18} r_H^{12}+11952 i q^8 r^{17} w
   r_H^{12}-4200 q^8 r^{16} w^2 r_H^{12}-34800 q^6
   r^{18} w^2 r_H^{12}\\&-5808 i q^6 r^{17} w^3 r_H^{12}+1932 q^6 r^{16} w^4
   r_H^{12}-6 i q^6 r^{15} w^5 r_H^{12}+17010 q^4
   r^{18} w^4 r_H^{12}\\&-5832 i q^4 r^{17} w^5 r_H^{12}-378 q^4 r^{16} w^6
   r_H^{12}-6966 q^2 r^{18} w^6 r_H^{12}+1458 i q^2
   r^{17} w^7 r_H^{12}\\& -3252 q^{10} r^{14} r_H^{14}-29433 q^8 r^{16}
   r_H^{14}-12720 i q^8 r^{15} w r_H^{14}+324
   r^{18} w^8 r_H^{12}\\&+3924 q^8 r^{14} w^2 r_H^{14}+27168 q^6 r^{16} w^2
   r_H^{14}+8088 i q^6 r^{15} w^3 r_H^{14}-1404 q^6
   r^{14} w^4 r_H^{14}\\&+6 i q^6 r^{13} w^5 r_H^{14}-5526 q^4 r^{16} w^4
   r_H^{14}+2664 i q^4 r^{15} w^5 r_H^{14}+180 q^4
   r^{14} w^6 r_H^{14}\\&-4560 q^{10} r^{14} \log (2) r_H^{14}+2106 q^2
   r^{16} w^6 r_H^{14}-594 i q^2 r^{15} w^7
   r_H^{14}-162 r^{16} w^8 r_H^{14}\\&-36480 q^8 r^{16} \log (2) r_H^{14}+4560 i q^8
   r^{15} w \log (2) r_H^{14}+3600 q^8 r^{14} w^2
   \log (2) r_H^{14}\\&+36480 q^6 r^{16} w^2 \log (2) r_H^{14}-8280 i q^6
   r^{15} w^3 \log (2) r_H^{14}-1080 q^6 r^{14} w^4
   \log (2) r_H^{14}\\&-720 i q^8 r^{15} w \log (4) r_H^{14}-11520 q^4
   r^{16} w^4 \log (2) r_H^{14}+900 i q^4 r^{15}
   w^5 \log (2) r_H^{14}\\&+540 q^8 r^{14} w^2 \log (4) r_H^{14}-216 i q^6
   r^{15} w^3 \log (32) r_H^{14}-90 q^6 r^{14} w^4
   \log (16) r_H^{14}\\&+240 i q^8 r^{15} w \log (64) r_H^{14}+84 q^8 r^{14}
   w^2 \log (1024) r_H^{14}+54 i q^4 r^{15} w^5
   \log (1024) r_H^{14}\\&-949 q^{10} r^{12} r_H^{16}-3129 q^8 r^{14}
   r_H^{16}-7376 i q^8 r^{13} w r_H^{16}+864 q^8
   r^{12} w^2 r_H^{16}+501 q^6 r^{14} w^2 r_H^{16}\\&+3828 i q^6 r^{13} w^3 r_H^{16}-219 q^6 r^{12} w^4
   r_H^{16}-2115 q^4 r^{14} w^4 r_H^{16}+594 i q^4
   r^{13} w^5 r_H^{16}\\&+949 q^{10} r^{10} r_H^{18}+4941 q^8 r^{12}
   r_H^{18}+7776 i q^8 r^{11} w r_H^{18}-798 q^8
   r^{10} w^2 r_H^{18}+1215 q^2 r^{14} w^6 r_H^{16}\\&-135 q^6 r^{12} w^2 r_H^{18}-4032 i q^6 r^{11} w^3
   r_H^{18}+159 q^6 r^{10} w^4 r_H^{18}+405 q^4
   r^{12} w^4 r_H^{18}\\&+1580 q^{10} r^{10} \log (2) r_H^{18}+18720 q^8
   r^{12} \log (2) r_H^{18}-306 i q^4 r^{11} w^5
   r_H^{18}-621 q^2 r^{12} w^6 r_H^{18}\\&+500 i q^8 r^{11} w \log (2) r_H^{18}-600 q^8 r^{10}
   w^2 \log (2) r_H^{18}-9600 q^6 r^{12} w^2 \log
   (2) r_H^{18}\\&+1080 i q^6 r^{11} w^3 \log (2) r_H^{18}+180 q^6
   r^{10} w^4 \log (2) r_H^{18}+1440 q^4 r^{12} w^4
   \log (2) r_H^{18}\\&+480 i q^8 r^{11} w \log (4) r_H^{18}-360 q^8 r^{10}
   w^2 \log (4) r_H^{18}+144 i q^6 r^{11} w^3 \log
   (32) r_H^{18}\\&+134 q^{10} r^8 r_H^{20}-3301 q^8 r^{10}
   r_H^{20}+1856 i q^8 r^9 w r_H^{20}-18 i q^4
   r^{11} w^5 \log (1024) r_H^{18}\\&-134 q^{10} r^6 r_H^{22}-66 q^8 r^8 w^2
   r_H^{20}+3144 q^6 r^{10} w^2 r_H^{20}-588 i q^6
   r^9 w^3 r_H^{20}-99 q^4 r^{10} w^4 r_H^{20}\\&+2529 q^8 r^8 r_H^{22}-1984 i q^8 r^7 w r_H^{22}+60
   q^8 r^6 w^2 r_H^{22}-2736 q^6 r^8 w^2
   r_H^{22}+576 i q^6 r^7 w^3 r_H^{22}\\&-280 q^{10} r^6 \log (2) r_H^{22}-4800 q^8 r^8 \log
   (2) r_H^{22}-840 i q^8 r^7 w \log (2)
   r_H^{22}+117 q^4 r^8 w^4 r_H^{22}\\&-80 i q^8 r^7 w \log (4) r_H^{22}+60 q^8 r^6 w^2
   \log (4) r_H^{22}+960 q^6 r^8 w^2 \log (2)
   r_H^{22}\\&-7 q^{10} r^4 r_H^{24}+1243 q^8 r^6 r_H^{24}-168 i
   q^8 r^5 w r_H^{24}-24 i q^6 r^7 w^3 \log (32)
   r_H^{22}-543 q^6 r^6 w^2 r_H^{24}\\&+7 q^{10} r^2 r_H^{26}+20 q^{10} r^2 \log (2)
   r_H^{26}-1127 q^8 r^4 r_H^{26}+184 i q^8 r^3 w
   r_H^{26}+477 q^6 r^4 w^2 r_H^{26}\\&+480 q^8 r^4 \log (2) r_H^{26}+140 i q^8 r^3 w \log
   (2) r_H^{26}-129 q^8 r^2 r_H^{28}+125 q^8
   r_H^{30}\biggr)\\&-q^2 \left(-4 r^4 r_H^4+r_H^8+3 r^8\right){}^2\biggl(-270 q^6 r^{16}+387 i q^6 r^{15} w+90 q^6 r^{14}
   w^2\\&+1152 q^4 r^{16} w^2-846 i q^4 r^{15} w^3-162 q^4
   r^{14} w^4-1134 q^2 r^{16} w^4+711 i q^2 r^{15}
   w^5
\end{split}
\end{equation}
\begin{equation}
\begin{split}
&+108 q^8 r^{10} r_H^4+1128 q^6 r^{12} r_H^4-1137 i
   q^6 r^{11} w r_H^4+72 q^2 r^{14} w^6+324 r^{16}
   w^6-216 i r^{15} w^7\\&-192 q^6 r^{10} w^2 r_H^4-3948 q^4 r^{12} w^2
   r_H^4+1032 i q^4 r^{11} w^3 r_H^4+78 q^4 r^{10}
   w^4 r_H^4+2592 q^2 r^{12} w^4 r_H^4\\&-24 q^8 r^6 r_H^8-1788 q^6 r^8 r_H^8+533 i q^6 r^7
   w r_H^8+6 q^6 r^6 w^2 r_H^8+45 i q^2 r^{11} w^5
   r_H^4+108 r^{12} w^6 r_H^4\\&-4 q^8 r^2 r_H^{12}+984 q^6 r^4 r_H^{12}+3408 q^4
   r^8 w^2 r_H^8-426 i q^4 r^7 w^3 r_H^8-1602 q^2
   r^8 w^4 r_H^8\\&+105 i q^6 r^3 w r_H^{12}-54 q^6 r_H^{16}-708 q^4
   r^4 w^2 r_H^{12}\biggr)\log{\left(1+\frac{r_{H}^2}{r^2}\right)}\Biggr\}
\end{split}
\end{equation}
{}
\end{document}